\documentclass{aastex62}
\pdfoutput=1 %for arXiv submission 
\usepackage{amsmath}
\usepackage{amstext}
\usepackage{bm}
\usepackage[figure,figure*]{hypcap}
\usepackage{url}
\usepackage{booktabs}
\usepackage{subfigure}
\usepackage{xspace}
\usepackage{soul}
\usepackage{threeparttable}
\usepackage{graphicx}
\usepackage[sort&compress]{natbib}
\usepackage{xcolor}

\defcitealias{brewer2018}{BF18}
\defcitealias{thorngren2016}{T16}

\shorttitle{Cool Giant Planet Host Star Abundances}
\shortauthors{Teske et al.}

\begin{document}

\title{Do Metal-Rich Stars Make Metal-Rich Planets?  New Insights on Giant Planet Formation from Host Star Abundances\footnote{This paper includes data gathered with the 6.5 m \textit{Magellan} Telescopes located at Las Campanas Observatory, Chile.}\footnote{Some of the data presented herein were obtained at the W. M. Keck Observatory, which is operated as a scientific partnership among the California Institute of Technology, the University of California and the National Aeronautics and Space Administration. The Observatory was made possible by the generous financial support of the W. M. Keck Foundation.}}

\correspondingauthor{Johanna Teske}
\email{jteske@carnegiescience.edu}
\author{Johanna K. Teske}
\altaffiliation{Hubble Fellow} 
\affiliation{Observatories of the Carnegie Institution for Science, 813 Santa Barbara Street, Pasadena, CA 91101, USA}
\author{Daniel Thorngren}
\affiliation{Department of Physics, University of California, Santa Cruz, USA}
\author{Jonathan J. Fortney}
\affiliation{Department of Astronomy and Astrophysics, University of California, Santa Cruz, USA}
\author{Natalie Hinkel}
\affiliation{Southwest Research Institute, 6220 Culebra Rd, San Antonio, TX 78238, USA}
\author{John M. Brewer}
\affiliation{Department of Astronomy, Yale University, 52 Hillhouse Avenue, New Haven, CT 06511, USA}
\affiliation{Department of Astronomy, Columbia University, 550 West 120th Street, New York, New York 10027, USA}
\affiliation{Dept. of Physics \& Astronomy, San Francisco State University, 1600 Holloway Ave., San Francisco, CA 94132, USA}

\begin{abstract}
The relationship between the compositions of giant planets and their host stars is of fundamental interest in understanding planet formation. The solar system giant planets are enhanced above solar composition in metals, both in their visible atmospheres and bulk compositions.  A key question is whether the metal enrichment of giant exoplanets is correlated with that of their host stars. Thorngren et al. (2016) showed that in cool ($T_{\rm{eq}}<1000$~K) giant exoplanets, the total heavy-element mass increases with total $M_p$ and the heavy element enrichment relative to the parent star decreases with total $M_p$. In their work, the host star metallicity was derived from literature [Fe/H] measurements. Here we conduct a more detailed and uniform study to determine whether different host star metals (C, O, Mg, Si, Fe, and Ni) correlate with the bulk metallicity of their planets, using correlation tests and Bayesian linear fits. We present new host star abundances of 19 cool giant planet systems, and combine these with existing host star data for a total of 22 cool giant planet systems (24 planets).  Surprisingly, we find no clear correlation between stellar metallicity and planetary residual metallicity (the relative amount of metal versus that expected from the planet mass alone), which is in conflict with common predictions from formation models. We also find a potential correlation between residual planet metals and stellar volatile-to-refractory element ratios. These results provide intriguing new relationships between giant planet and host star compositions for future modeling studies of planet formation.

\end{abstract}

\section{Introduction \label{sec:intro}} 
Since the early days of exoplanet detection, there has been interest in comparing the properties of host stars and their planets, particularly their compositions. From many studies \citep[e.g.,][]{gonzalez1997,santos2004,fischer&valenti2005,johnson2010,mortier2013}, we now know that the occurrence rate of giant close-in planets is enhanced around stars with higher metallicity ([Fe/H]\footnote{[X/H]=log($N_{\rm{X}}$/$N_{\rm{H}}$)$_{\rm{star}}$ - log($N_{\rm{X}}$/$N_{\rm{H}}$)$_{\rm{solar}}$ = $A$(X)$_{\rm{star}}$-$A$(X)$_{\rm{solar}}$} or [$m$/H]), while the occurrence rate enhancement seems to decrease, if not completely disappear, with decreasing planet mass and radius (\citealt{buchhave2014}; \citealt{buchhave2015}; \citealt{schlaufman2015}; \citealt{wang&fischer2015}; \citealt{winn2017}; \citealt{petigura2018}; but see also \citealt{mulders2016}; \citealt{zhu2016};  \citealt{wilson2018}). This trend is important because it informs our picture of planet formation, indeed pointing toward core accretion as a dominant mechanism -- more metal-rich stars are indicative of more metal- or solid-rich protoplanetary disks, which enable giant planet cores of $\sim$10~$M_{\oplus}$ to grow from planetesimals and then accrete gas more quickly before the disk disperses \citep[e.g.,][]{pollack1996,mordasini2009}.

Beyond just a broad picture of core accretion, however, we would like to verify a more detailed model of giant planet formation that incorporates factors like planet or planetesimal migration through a compositionally varying disk \citep[e.g.,][]{alibert2005, raymond2006, alidib2014}, the relative importance of planetesimal versus pebble versus gas accretion \citep{booth2017,hasegawa2018}, different opacities of accreted material \citep[e.g.,][]{mordasini2014b}, and mixing/composition gradients within planets \citep[e.g.][]{vazan2016,helled2017}, to name a few. Verification of more detailed models requires more observational constraints, both in the number of observed planets and their parameters. The goal of comparing such models to observations is to understand when, how, and from which material giant planets formed, both in the solar system and beyond. 

In particular, studying the heavy-element enrichment of giant planet interiors \citep{guillot2006,burrows2007,miller&fortney2011,thorngren2016} and their atmospheres \citep{madhusudhan2011,fortney2013,konopacky2013,kreidberg2014,wakeford2017} has emerged as a potentially productive path forward. As shown in \citet{mordasini2016}, comparing different population-synthesis models to observations of solar system and giant exoplanet interior and atmospheric compositions can place constraints on the level of planetesimal enrichment (they find it must be the dominant source of heavy elements). In Mordasini et al. as well as in \citet{miller&fortney2011} and \citeauthor{thorngren2016} (2016; described in more detail below), heavy-element enrichment of planets is considered relative to the metallicity of their host stars. However, as suggested by Miller \& Fortney and \citetalias{thorngren2016}, examining trends between planet composition and more detailed stellar abundances (separating out  metallicity into Fe, Mg, Si, Ni, O, and C) may reveal new insights into planet formation, which helps refine the sources of planetary heavy-element enrichment. In this paper we follow this suggestion and look for trends between the bulk interior composition of cool, giant planets and different host star abundances. 

\subsection{Summary of T16}
The giant planet heavy-element abundances used in this paper come from the modeling formalism described in \citet[][\citetalias{thorngren2016}]{thorngren2016}. In this work, the authors investigated a sample of cool ($T_{\rm{eq}} \lesssim$~1000 K, or $F_{\rm{incident}} \lesssim$ 2$\times10^{8}$ erg~s$^{-1}$) giant (20 $M_{\oplus} < M_{p} < $ 20 $M_{J}$), transiting planets with relative uncertainties in mass and radius both smaller than $50\%$.  The planets they selected had much smaller uncertainties, averaging $10\%$ in mass and $20\%$ in radius. The authors created one-dimensional structure models for the planets by solving the equations of state under hydrostatic equilibrium, assuming a core composed of a 50/50 rock-ice mixture and a homogeneous convective envelope composed of a mixture of H/He-rock-ice.  Heat flow out of the planet was mediated by a radiative atmosphere (the upper boundary).  The authors drew 10,000 samples from the probability distributions given by the uncertainties in mass, radius, and age of the observed planets; the inferred heavy-element mass for each draw was the amount that reproduced the sampled radius.  The resulting planetary heavy-element mass ($M_z$) distributions were single peaked and roughly Gaussian, with typical errors in $M_z$ between 10 and 30\%. Changes in $M_z$ due to uncertainties in the exact distribution of heavy elements between the core and envelope, and the equations of state, were smaller than those from uncertainties in planetary mass and radius.  As such, their approach did not require the assumption of a particular formation mechanism, such as core accretion or gravitational instability.  The authors also collected measured [Fe/H] host star values from the literature, which tended to have fairly high uncertainties and were drawn from different sources.  The upshot was that giant planet bulk metallicities could be obtained from structural evolution models.

Together with their planet metallicities, \citetalias{thorngren2016} used literature values of the metallicities of each planet-hosting star to examine the correlations between $M_p$, $M_z$, $Z_{\rm{planet}}$ (metal mass fraction), and the relative planet-to-star metal enrichment, $Z_{\rm{planet}}/Z_{\rm{star}}$. They found a very clear positive correlation between the predicted heavy-element mass in cool giant planets versus their total mass,  which is in line with the core-accretion model.  However, the authors observed no strong trends between heavy-element mass and host star [Fe/H], nor even between total mass and host star [Fe/H]\footnote{In addition, the most massive planets being are found far less often around low-metallicity stars}.%, which is surprising given the expected extension of the metal content in stars to the protoplanetary disks, out of which planets form. 

However, \citetalias{thorngren2016} did observe a trend between $Z_{\rm{planet}}/Z_{\rm{star}}$ and total planet mass $M_p$, as originally suggested by \cite{miller&fortney2011} and then matched in population-synthesis models \citep{Mordasini2014a}. The $Z_{\rm{planet}}/Z_{\rm{star}}$ trend is stronger than considering $Z_{\rm{planet}}$ alone versus $M_p$, suggesting that the host star metallicity does have some influence on giant planet mass and thus heavy-element content. Interestingly, even the most massive planets in the \citetalias{thorngren2016} sample have $Z_{\rm{planet}}/Z_{\rm{star}} > 1$, indicating that their envelopes are likely also enriched in metals (because their cores cannot account for the entire $M_z$).  \citetalias{thorngren2016} showed that their fit to $Z_{\rm{planet}}/Z_{\rm{star}}$ vs. $M_p$ is wellmatched by a simple core-accretion planet formation model in which the planet metallicity is determined by accretion of solids from the gravitational zone of influence in the disk around the planet.  

\subsection{This Work}
In this work, we take the next important step on the composition path by conducting a similar study to that of \citetalias{thorngren2016} across a range of stellar abundances and thus $Z_{\rm{star}}$ variations. This work relies on a multi-year spectroscopic survey of cool giant planet host stars to measure each star's C, O, Si, Mg, Ni, and Fe abundances, and an updated Bayesian analysis to search for trends between planet the heavy-element content of the planet and composition of the host star, taking into account correlations between stellar abundances due to Galactic chemical evolution. In \S\ref{sec:obs} we describe the host star observations and abundance measurements in detail; for many of the host stars in our sample, only a bulk [$m$/H] or [Fe/H] was measured, so this represents the first measurement of their more detailed abundances. We attempt to include sufficient detail to enable easy reproduction of our results. In \S\ref{sec:obs} we also include a comparison with the abundance work of \cite{brewer2018} for a handful of overlapping targets, with which we find good agreement. In \S\ref{sec:analysis} we describe our updated comparison of the planet and host star compositions, again building on the work in \citetalias{thorngren2016}.  We include a discussion and summary of our results in \S\ref{sec:discussion}, where we invite planet formation theorists to consider the (lack of) trends we find in their future models.

\section{Observations and Stellar Abundance Measurements} \label{sec:obs}
\subsection{Sample Selection}
Our list of candidate stars hosting relatively cool transiting giant planets was constructed with the same selection criterion as used in \cite{thorngren2016}: $T_{\rm{eq}}<$ 1000~K (incident flux $<2 \times 10^{8}$ erg~s$^{-1}$~cm$^{-2}$), $20~M_\oplus < M < 20~M_J$, and $M_p$ and $R_p$ measured with errors $\leq50$\%. From these, the systems chosen for this study are biased toward the brightest host stars to allow for high-resolution and spectroscopy with high signal-to-noise ratio (S/N). Similar to the sample in \citetalias{thorngren2016}, the $M_p$ errors for our sample are typically around $\sim$10\%, but as high as 30\%, and the $R_p$ errors for our sample are typically around $\sim$5\%, but as high as 25\%.\footnote{Based on accessing the NASA Exoplanet Archive on 2019 June 3.} The lower $M_P$ limit of our sample is similar to that in \citetalias{thorngren2016} ($\sim$20 $M_{\oplus}$), but the upper limit differs somewhat because so few planets are in the multiple-Jupiter mass range; our most massive planet has 5.4~$M_J$, compared to 10.1~$M_J$ in \citetalias{thorngren2016}. In the interior structure modeling from \citetalias{thorngren2016}, this mass is between the lower bound where the distinction between rock and ice becomes important (around the mass of Neptune), and the upper bound at the hydrogen-burning limit (deuterium fusion only really changes the initial conditions). Different formation mechanisms might affect the initial composition and entropy and structure of the planet, but composition is directly estimated by our model, the initial entropy is quickly ``forgotten'' (e.g., \citealt{marley2007}), and the exact structure is a second-order effect (discussed in \citetalias{thorngren2016}). Thus, the models of \citetalias{thorngren2016} (and this paper) are agnostic about the formation mechanism.

\subsection{Details of Observations}
The results presented here are based on the analysis of observations from three echelle spectrographs: the High Resolution Echelle Spectrometer (HIRES; \citealt{Vogt1994}) on the 10 m Keck I telescope, the \textit{Magellan} Inamori Kyocera Echelle (MIKE) spectrograph \citep{bernstein2003} on the 6.5 m Magellan II Telescope, and the High Dispersion Spectrograph (HDS; \citealt{Noguchi2002}) on the 8.2 m Subaru Telescope. All observations are logged in Table \ref{tab:obslog}, and the instrument configurations are detailed in Table \ref{tab:platform}. To facilitate a differential abundance
analysis of the host stars with respect to the Sun (as indicated by the bracket notation), spectra of the Sun as reflected moonlight were taken with Subaru/HDS, as reflected light from the asteroids Iris and Vesta with \textit{Magellan} II/MIKE, and as reflected light from Vesta with Keck I/HIRES (PID N014Hr; PI Marcy, third row of Table \ref{tab:platform}).

\subsubsection{Keck I/HIRES}
The Keck I data analyzed here were obtained through programs N026Hr (PI Teske), and U042Hr, U049Hr, U172Hr, and U097 (PI Fortney), between 2014 and 2017. The details of these observations are listed in the second row of Table \ref{tab:platform}. We reduced and extracted wavelength-calibrated stellar spectra from the raw data frames using the MAKEE pipeline\footnote{www.astro.caltech.edu/~tb/makee/} with corresponding bias ($\sim$3), flat ($\sim$30), ThAr ($\sim$2-4), and trace star (bright B star or the target star itself) frames for each target frame separately. The frames were then combined in IRAF\footnote{IRAF is distributed by the National Optical Astronomy Observatory, which is operated by the Association of Universities for Research in Astronomy (AURA) under cooperative agreement with the National Science Foundation.} with the \texttt{scombine} task and normalized by fitting a several-order spline function. 

\subsubsection{\rm{Magellan} \textit{II/MIKE}}
The \textit{Magellan} II data analyzed here were obtained through Carnegie time (PI Teske) between 2014 and 2017. MIKE consists of two arms separated by a dichroic, with the standard grating angles covering 3350-5000~{\AA} in the blue and 4900-9300~{\AA} in the red. Further details of these observations are listed in the fourth row of Table \ref{tab:platform}. The data were reduced using the Carnegie Python Distribution (CarPy) MIKE pipeline \citep{Kelson2000,Kelson2003}, typically including the following calibration frames: $\sim$15-20 quartz flats taken with an internal incandescent lamp for tracing the orders, 25 ``milky'' quartz flats taken through a diffuser to provide good pixel-to-pixel sensitivity correction, both with the incandescent lamp for the red arm and with a hot O or B star for the blue arm, and a few ThAr frames for wavelength calibration. The bias level was calculated by using the overscan regions at the end of each column/row. The frames were combined in the pipeline to produce a multispec file with, for every order, the sum of the sky over the extraction aperture, the sum of the object over the extraction aperture, the expected noise from sum of object and sky plus the read noise, the S/N spectrum (per pixel), the sum of the lamp spectrum over the extraction aperture, the blaze function, and the spectra divided (flattened) by the blaze.

\subsubsection{Subaru/HDS}
The only star in our sample with Subaru/HDS observations is HD 80606, originally published in \cite{teske2014}. We refer the reader to that paper for details of Subaru/HDS reductions; the important details of the setup are listed in the fifth row of Table~\ref{tab:platform}.  

\begin{deluxetable}{lcccccc}
    \rotate
    \tablecolumns{7}
    \tablewidth{0pc}
    \tabletypesize{\scriptsize}
    \tablecaption{Observing Log \label{tab:obslog}}
    \tablehead{
        \colhead{Star} & \colhead{$V$ mag} & \colhead{Date} & \colhead{Exposures} & \colhead{T$_{\rm{exp}}$} & \colhead{Approximate S/N of}  & \colhead{Platform} \\
        \colhead{ } & \colhead{ } & \colhead{(UT)} & \colhead{ }& \colhead{(s)} & \colhead{Combined Frames (at 6300\,\AA)} & {}}
    \startdata
        Vesta	&		&	2006 Apr 16	&	3	&	216, 232, 241	&	470	&	Keck I/HIRES archive	\\
        Vesta	&		&	2014 Jul 12	&	3	&	1500	&	200	&	Magellan II /MIKE	\\
        Moon	&		&	2012 Feb 10	&	2	&	1, 5	&	590	&	Subaru/HDS	\\
        Iris	&		&	2015 Apr 30	&	3	&	200, 600, 800	&	230	&	Magellan II/MIKE	\\
        \hline
        HD80606	&	9	&	2012 Feb 11	&	2	&	600	&	290	&	Subaru/HDS	\\
        WASP-8	&	9.8	&	2014 Jul 12	&	6	&	1200	&	365	&		Magellan II/MIKE\\
        HAT-P-15	&	12.4	&	2014 Sep 12	&	2	&	1387, 1492	&	220	&	Keck I/HIRES	\\
        HAT-P-17	&	10.4	&	2014 Aug 13	&	2	&	350, 816	&	260	&	Keck I/HIRES	\\
        Kepler-89	&	12.4	&	2014 Aug 26	&	1	&	3034	&	190	&	Keck I/HIRES	\\
        Kepler-419	&	14	&	2015 Sep 24	&	5, 1	&	1800 (5), 1200 (1)	&	200	&	Keck/HIRES	\\
        Kepler-432	&	13	&	2015 Sep 24	&	4	&	1800	&	280	&	Keck I/HIRES	\\
        WASP-84	&	9.4	&	2015 Apr 29	&	3	&	750	&	260	&	Magellan II/MIKE	\\
        WASP-139	&	12.4	&	2016 Sep 20	&	8	&	1500	&	220	&	Magellan II /MIKE	\\
        K2-24	&	11.3	&	2016 Jul 10	&	3	&	1200	&	190	&	Magellan II/MIKE	\\
        WASP-130	&	11.1	&	2016 Aug 13	&	4	&	1200	&	300	&	Magellan II/MIKE	\\
        WASP-29	&	11.3	&	2016 Sep 19	&	5	&	1320	&	290	&	Magellan II/MIKE	\\
        Kepler-9	&	13.9	&	2010 May 26	&	1	&	180	&		&	Keck I/HIRES	\\
        Kepler-9	&		&	2010 Jun 3, 25, 26, 30	&	3, 1	&	2700 (1), 300 (3)	&		&	Keck I/HIRES \\
        Kepler-9	&		&	2016 Jul 16, 17, 18, 22, 31	&	3,1,1	&	 1800 (3), 1768 (1), 900 (1)	&		&	Keck I/HIRES	\\
        Kepler-9	&		&	2016 Aug 12, 13, 15, 16, 17, 18, 20, 21	&	8, 1	&	900 (8), 1800 (1)	&		&	Keck I/HIRES	\\
        Kepler-9	&		&	2016 Sep 20, 21 22	&	5	&	900	&		&	Keck I/HIRES	\\
        Kepler-9 combined	&		&		&		&		&	250	&		\\
        K2-19	&	13	&	2017 Jan 21; 2017 Feb 4, 8, 9, 22	&	10	&	900	&	240	&	Keck I/HIRES	\\
        K2-27	&	12.6	&	2017 Jan 21, 22; 2017 Feb 4, 9	&	6, 1	&	900 (6), 800 (1)	&	240	&	Keck I/HIRES	\\
        K2-139 (EPIC 218916923)	&	11.7	&	2017 May 15	&	7	&	1200	&	220	&	Magellan II/MIKE	\\
        Kepler-145	&	12.1	&	2017 Aug 7, 08	&	1, 8	&	600, 1200	&	280	&	Keck I/HIRES	\\
        Kepler-277	&	13.6	&	2017 Aug 8	&	6	&	1320	&	100	&	Keck I/HIRES	\\
        CoRoT-9	&	13.5	&	2017 Aug 8	&	11	&	1200	&	190	&	Keck I/HIRES	\\
       % Kepler-128	&	11.7	&	2017 Aug 05	&	6	&	960	&	350	&	Keck I/HIRES	\\
        Kepler-539	&	12.6	&	2017 Aug 5	&	7	&	1080 (5), 1023 (1), 1500 (1)	&	220	&	Keck I/HIRES	\\
    \enddata
\end{deluxetable}

\begin{deluxetable}{cccc}
    \tablecolumns{4}
    \tablewidth{0pc}
    \tabletypesize{\normalsize}
    \tablecaption{Observing Platform Details \label{tab:platform}}
    \tablehead{\colhead{Platform} & \colhead{Slit} & \colhead{$R$} &  \colhead{Wavelength Coverage}\\
        \colhead{ } & \colhead{(and filter, if applicable)} & \colhead{($\frac{\lambda}{\Delta\lambda}$)} &\colhead{(\AA)}}
    \startdata
        Subaru/HDS	&	0.6"	&	60000	&	$\sim$4450-5660; 5860-7100$^{a}$	\\
        Keck I/HIRES	&	C1, C2 (0.86"); kv370+clear	&	48000, 72000	&	$\sim$4380-8800	\\
        &	&	&	(varies, some spectra have 5000-8000)	\\
        Keck I/HIRES archive	&	E3 (0.4"); kv370$+$clear &	103000	&	~3680-7900	\\
        Magellan II/MIKE	&	0.35", 0.5"	&	65000, 45000	&	$\sim$3350-5000, 4900-9500$^{a}$	\\
    \enddata
    \tablenotetext{a}{Wavelength coverage across two separate CCDs.}
\end{deluxetable}

\subsection{Stellar Parameters}
The stellar parameters -- effective temperature ($T_\mathrm{eff}$), log~$g$, microturbulent velocity ($\xi$), and metallicity ([Fe/H]) -- were determined using a line-by-line differential (with respect to the Sun) equivalent width (EW) analysis similar to that reported by \cite{teske2014}. In each spectrum (including the solar reference), we measured the absorption strengths of Fe I and II lines using the IRAF \texttt{splot} task and a Gaussian profile, with line parameters from the Vienna Atomic Line Database (VALD; \citealt{piskunov1995,ryabchikova1997,kupka1999,kupka2000,ryabchikova2015}). Typically, we used $\sim$70 Fe I and $\sim$12 Fe II lines, but the number and specific wavelengths of the lines varied depending on the wavelength coverage and strength of the lines in each star, see Table \ref{tab:lines}. 

The lines were drawn from \cite{ramirez2014} and \cite{teske2014}, the latter of which was based on \cite{schuler2011}, which was drawn from \cite{thevenin1990}. Because we did not collect all of the host star spectra at once and were continually analyzing new observations while trying to maintain a differential analysis, we grouped stars by collection date and instrument, and for each group compared the solar reference spectrum and host star spectra line by line. The [Fe/H] values were determined differentially ($A$(Fe)$_{\rm{star}}$ - $A$(Fe)$_{\rm{Sun}}$) on a line-by-line basis, and only the lines that were measurable in the target star spectrum were used to calculate its respective [Fe I/H] and [Fe II/H] values. This resulted in multiple sets of solar measurements, with small differences in absolute values but measured in exactly the same way as that of each partner star(s), preserving the same continuum region choices. Such a strictly differential analysis is important for deriving the most precise and accurate relative stellar abundances \citep[e.g.,][]{bedell2014}.  

Given the EW line measurements, the best-fit stellar parameters result from fulfilling ionization equilibrium (Fe I and Fe II lines give the same [Fe/H] abundance, averaged across all lines) and excitation equilibrium (no correlation between the [Fe I/H] values and the lower excitation potential of the lines, $\chi$), as well as requiring [Fe I/H] values to show no correlation with the ``reduced'' EW values (log(EW/$\lambda$)). Initial guesses at $T_{\rm{eff}}$, log~$g$, $\xi$, and [Fe/H] for each host star were taken from the literature and then iterated on by slightly changing each parameter until a global solution was reached that fulfilled the criteria above. Throughout the analysis, the solar parameters were held fixed at $T_{\rm{eff}}$=5777 K, log $g=$4.44 dex, [Fe/H]=0, and $\xi=$1.38 km~s$^{-1}$.  

The EW measurements were translated into elemental abundances using the curve-of-growth method via the 2014 or 2017 version\footnote{There are no significant differences between the 2014 and 2017 MOOG versions that affect the abundance determinations in this work.} of the spectral analysis code MOOG \citep{sneden1973}, specifically, the \texttt{abfind} driver, and ``standard composition''  Kurucz ATLAS9 ``odfnew'' stellar atmosphere models{\footnote{See http://kurucz.harvard.edu/grids.html} \textbf(for the ``by-hand'' method) or MARCS 1D-LTE stellar atmosphere models \citep[][for the rest of the stars]{gustafsson2008} linearly interpolated to the appropriate stellar parameters; both models provide similar results for the type of stars in our sample.} For five of the stars (WASP-8, HAT-P-15, HAT-P-17, Kepler-89, and HD 80606) we adopted the same by-hand methodology as \cite{teske2014}, based on finding the parameters that resulted in no correlations (significantly small $r$ correlation coefficients) between [Fe I/H] and $\chi$ values, no correlations between [Fe I/H] and reduced EW values, and the same [Fe/H] from both Fe I and II lines. With this method, stellar parameter error estimation involves determining the changes in parameters needed to cause 1$\sigma$ correlations between [Fe I/H] and $\chi$ (for $T_{\rm{eff}}$), and between [Fe I/H] and reduced EW (for $\xi$), and uses a iterative scheme based on the differences in [Fe I/H] and [Fe II/H] for log~$g$ (detailed in \citealt{teske2013}). 

For the rest of the stars, we used the new publicly available \texttt{Qoyllur-quipu} \citep[$q^2$,][]{ramirez2014} Python package\footnote{https://github.com/astroChasqui/q2}, a MOOG wrapper that uses EWs as input to derive stellar parameters and abundances. Instead of using correlation coefficients to determine when the stellar parameters satisfy ionization and excitation equilibrium, $q^2$ automatically determines the conditions that result in a slope consistent with zero between [Fe I/H] and $\chi$ and between [Fe I/H] and reduced EW. The error estimation for $T_{\rm{eff}}$, log~$g$, and $\xi$ in $q^2$ is described in \cite{ramirez2014} and is based on \cite{epstein2010} and \cite{bensby2014}; these errors are added in quadrature with the standard error of the mean in the line-to-line [Fe/H] abundance to estimate the error in [Fe/H]. Given the measured EWs in Table \ref{tab:lines}, the results presented here can be fully reproduced with $q^2$.

\begin{figure}[htp]
    \centering
    \begin{minipage}{.47\linewidth}
        \centering
	    \includegraphics[width=\textwidth,clip]{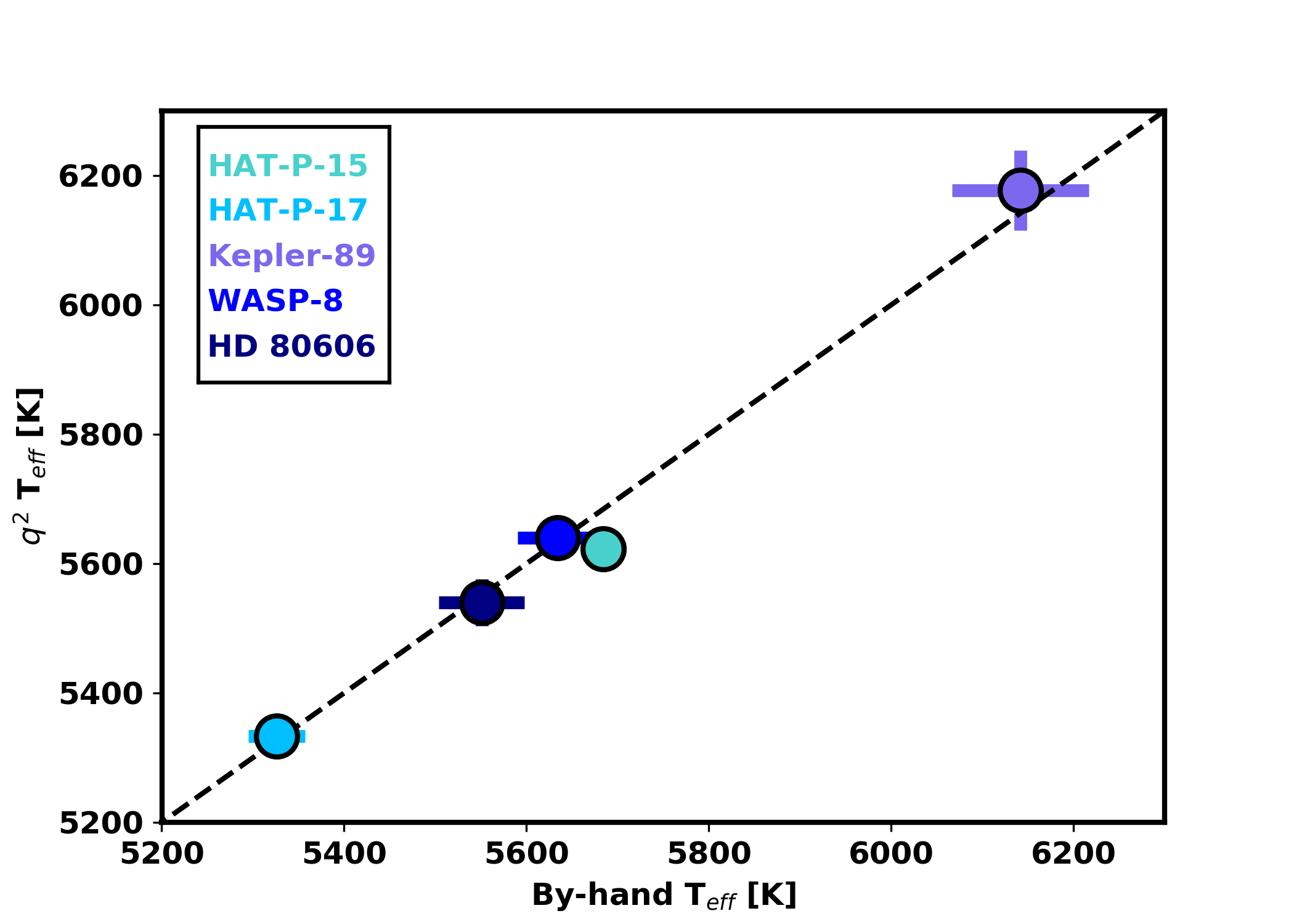} 
    \end{minipage}
    \begin{minipage}{0.47\linewidth}
        \centering
	    \includegraphics[width=\textwidth,clip]{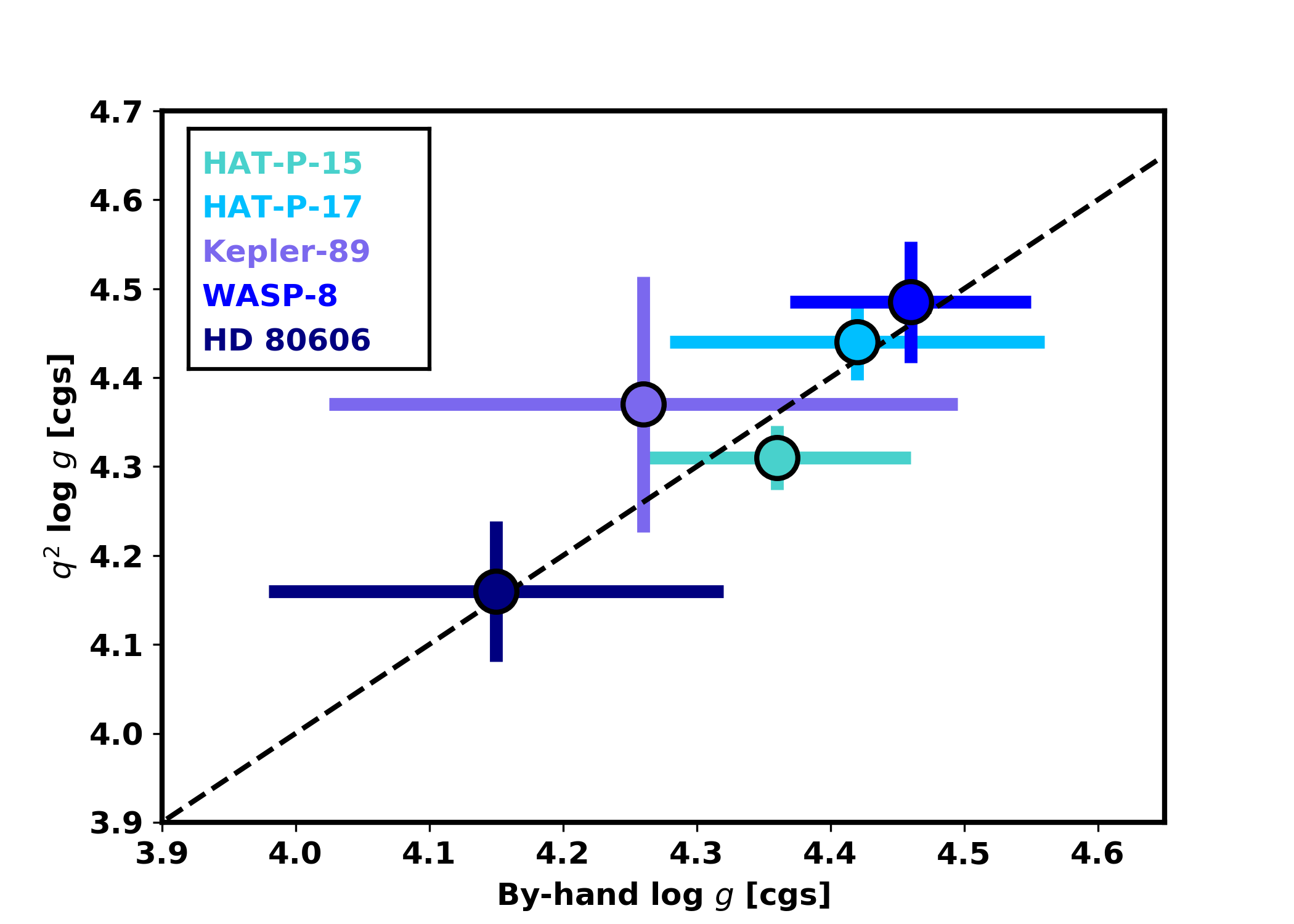} 
    \end{minipage}
    \begin{minipage}{0.47\linewidth}
        \centering
	    \includegraphics[width=\textwidth,clip]{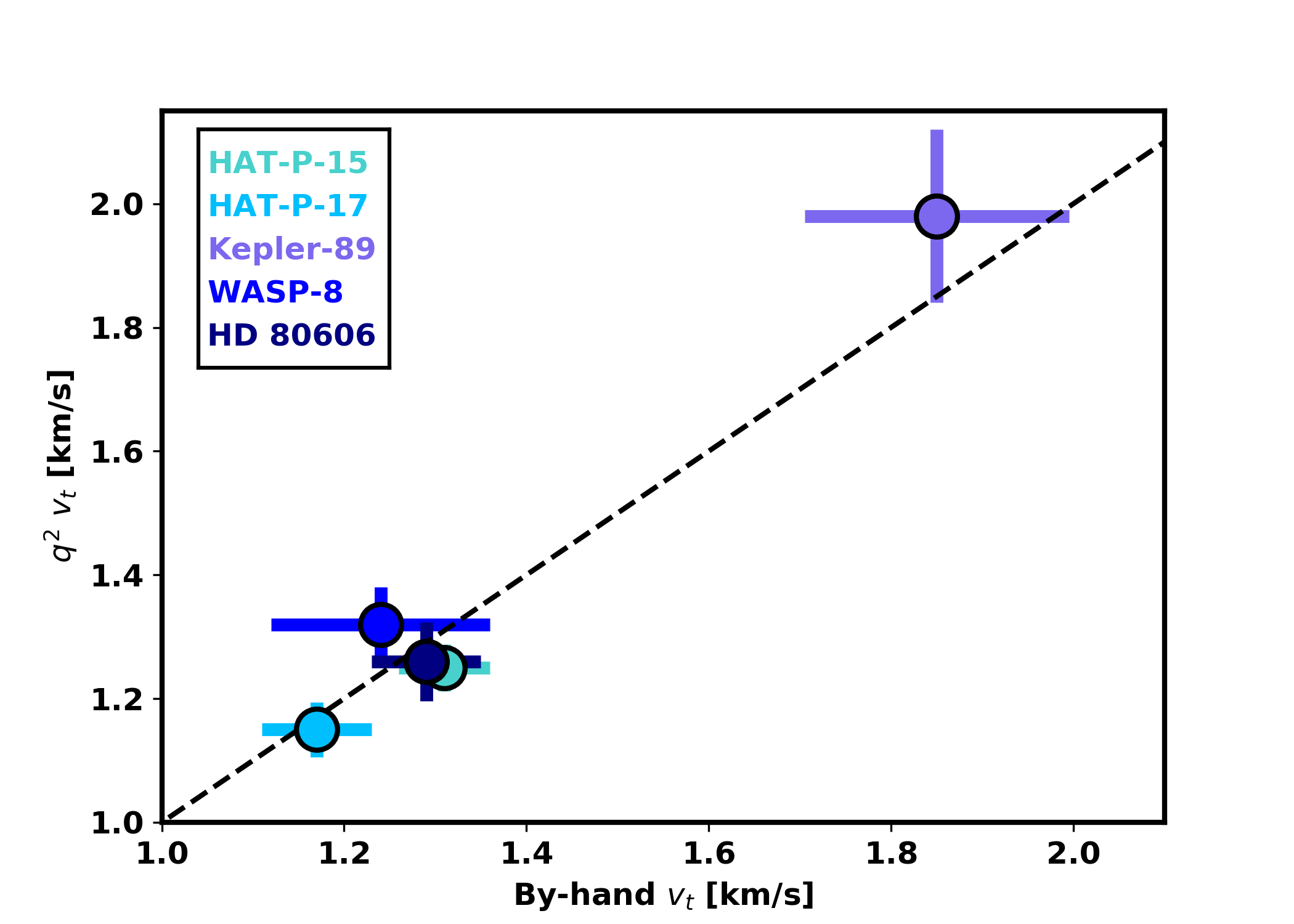} 
    \end{minipage}
    \begin{minipage}{0.47\linewidth}
        \centering
	    \includegraphics[width=\textwidth,clip]{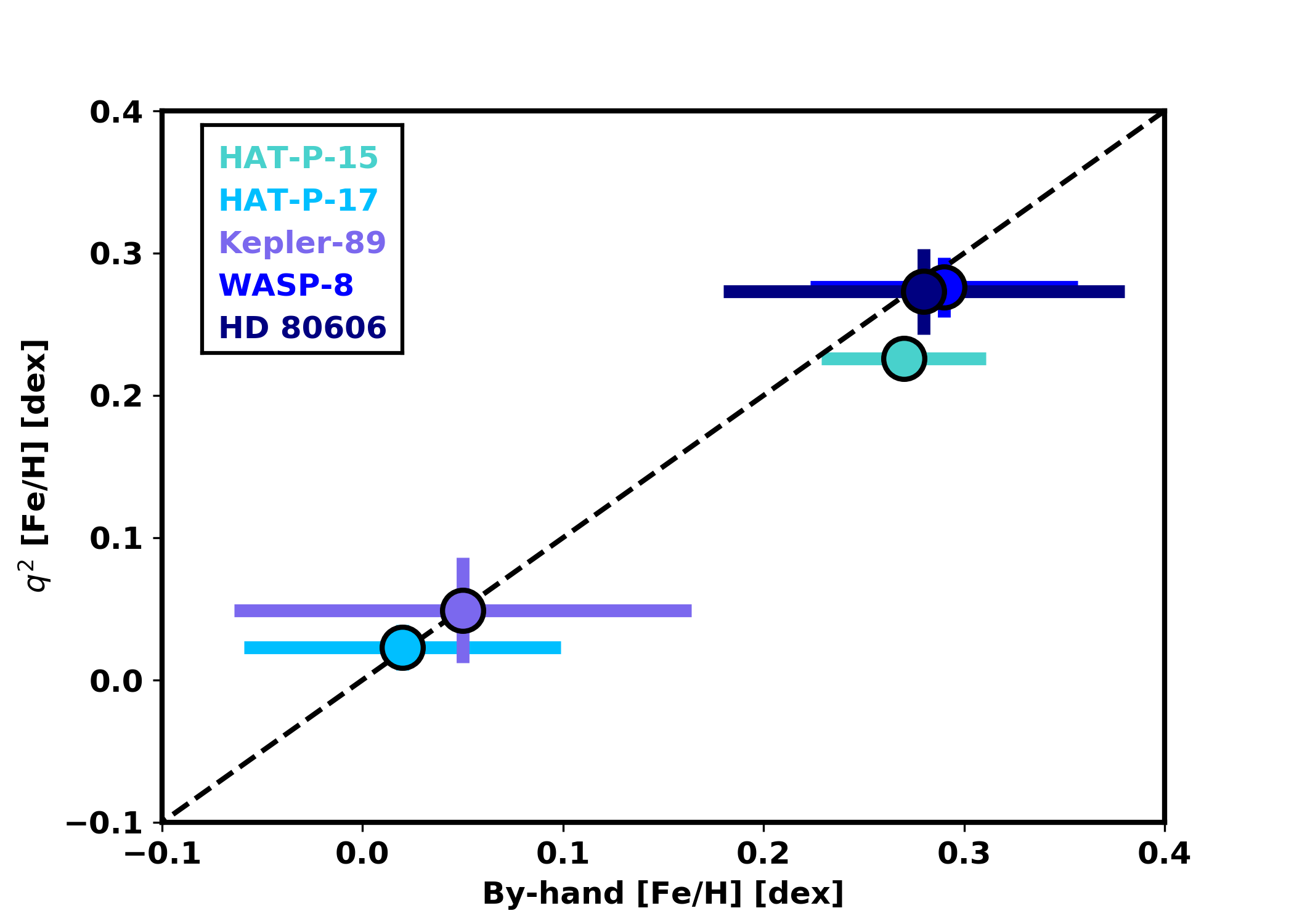} 
    \end{minipage}
    \caption{Comparisons of stellar parameters for five host stars in our sample, derived from a non-automated, by-hand method ($x$-axis) versus the automated $q^2$ method ($y$-axis), both using the same EW input measurements and spectral analysis code (MOOG). Dashed lines represent a 1:1 correlation. As expected, the parameters overlap within the errors.}
    \label{fig:handvq2}
\end{figure}

Our final stellar parameters are listed in Table \ref{tab:params}. As a consistency check, we rederived the stellar parameters of WASP-8, HAT-P-15, HAT-P-17, Kepler-89, and HD 80606 using $q^2$ instead of our completely by-hand analysis. The methods are in principle very similar, and because we use the same input EW measurements, we expect the parameters to be nearly identical. We find exactly this, as shown in Figure \ref{fig:handvq2}, where each different colored point represents a different star, and we show each of the stellar parameters in a separate subplot. The dashed line in each subplot represents a perfect correlation between by-hand and $q^2$ parameters. We are thus confident that adopting our original by-hand results for these stars does not introduce any additional offset, and do so moving forward.

\begin{deluxetable}{lccccccccc}
    \tablecolumns{10}
    \tablewidth{0pc}
    \tabletypesize{\scriptsize}
    \tablecaption{Measured Lines and Equivalent Widths \label{tab:lines}}
    \tablehead{
        \colhead{Ion} & \colhead{$\lambda$} & \colhead{$\chi$} &\colhead{log $gf$} & \colhead{Vesta-HIRES} & \colhead{HAT-P-15} & \colhead{HAT-P-17} & \colhead{Kepler-89} & \colhead{Vesta-MIKE} & \colhead{WASP-8}\\
        \colhead{} & \colhead{({\AA})} & \colhead{(ev)} & \colhead{dex} & \colhead{EW (m{\AA})} & \colhead{EW (m{\AA})} & \colhead{EW (m{\AA})}  & \colhead{EW (m{\AA})} & \colhead{EW (m{\AA})} & \colhead{EW (m{\AA})}}
    \startdata
        C I	&5052.167	   &7.685	&-1.304	&	33.3	&33.5	&20.3	&		&33.4	&30.2 \\
        C I	&5380.337	   &7.685	&-1.615	&	20.9	&26.5	&15.0	&40.0   &19.4	&22.2\\
        C I &6587.61	   &8.537	&-1.021	&	13.8   &15.4    &8.6	&14.6   &13.9	&14.7\\
        C I	&7111.47	   &8.64	&-1.074	&	10.4	&15.9	&       &10.4   &10.8	&12\\
        \big[O I\big] &6300 blends	&0.000 &-9.717	&	5.0	    &7.2	&6.5    &2.7	&5.4    &8.2\\			
        O I	&7771.944      &9.146	&0.37  &	69.7&	70.5    &41.2	&71.5	&71.3   &62.7\\
        O I &7774.166	  &9.146&	0.22  &		63.2&	62.4	&34.0 	&65.9	&61.1	&59.9\\
        O I	&7775.388	  &	9.146&	0.000  &	44.8&	45.8	&23.9	&44.8	&47.7&	45.4\\
    \enddata
    \tablecomments{This table is available in its entirety in a machine-readable form in the online journal. A portion is shown here for guidance regarding its form and content.}
\end{deluxetable}

\begin{deluxetable}{lcccccccc}
    \tablecolumns{9}
    \tablewidth{0pc}
    \tabletypesize{\scriptsize}
    \tablecaption{Stellar Parameters \label{tab:params}}
    \tablehead{  \colhead{Star} & \colhead{T$_\mathrm{eff}$ (K)} & \colhead{T$_\mathrm{eff}$ error (K)} & \colhead{log~$g$ (dex)} & \colhead{log~$g$ error (dex)} & \colhead{$\xi$ (km~s$^{-1}$)}  & \colhead{$\xi$ error (km~s$^{-1}$)} & \colhead{[Fe/H] (dex)} & \colhead{[Fe/H] error (dex)}}
    \startdata
        WASP-8 		&5634	&44	&4.46	&0.09	&1.24	&0.12 &0.29 & 0.07 \\
        HAT-P-15 	&5684	&25	&4.36	&0.10	&1.31	&0.05 &0.27 & 0.04 \\
        HD80606		&5551	&47	&4.15	&0.17	&1.29	&0.06 &0.28 & 0.10 \\
        HAT-P-17	&5326	&31	&4.42	&0.14	&1.17	&0.06 &0.02 & 0.08 \\
        Kepler-89	&6142	&75	&4.26	&0.24	&1.85	&0.15 &0.05 & 0.11 \\
        Kepler-419	&6214	&70	&3.90	&0.23	&2.34	&0.23 &0.03 & 0.05 \\
        Kepler-432	&5078	&40	&3.55	&0.08	&1.37	&0.05 &0.02 & 0.03 \\
        WASP-84		&5350	&31	&4.63	&0.06	&1.46	&0.07 &0.05 & 0.02 \\
        WASP-139	&5239	&16	&4.49	&0.04	&1.22	&0.05 &0.09 & 0.02 \\
        K2-24	    &5696	&14	&4.34	&0.04	&1.39	&0.04 &0.37 & 0.02 \\
        WASP-130	&5656	&11	&4.39	&0.03	&1.31	&0.03 &0.29 & 0.01 \\
        WASP-29		&5006	&63	&4.53	&0.08	&1.61	&0.14 &0.12 & 0.05 \\
        Kepler-9	&5778	&10	&4.47	&0.03	&1.36	&0.03 &0.06 & 0.01 \\
        K2-19		&5418	&21	&4.54	&0.04	&1.40	&0.05 &0.03 & 0.02 \\
        K2-27		&5342	&25	&4.53	&0.04	&1.25	&0.05 &0.11 & 0.02 \\
        K2-139 (EPIC 218916923)	&5342	&22		&4.51	&0.04	&1.380	&0.050 &0.18 & 0.02 \\
        Kepler-145	&6078	&25	&4.01	&0.05	&1.91	&0.05 &-0.03 & 0.02 \\
        Kepler-277	&6083	&29	&4.13	&0.06	&1.84	&0.06 &0.04 & 0.02 \\
        CoRoT-9		&5646	&10	&4.47	&0.03	&1.28	&0.03 &-0.02 & 0.01 \\
       % Kepler-128	&6087	&24	&4.10	&0.05	&1.95	&0.06 &-0.11 & 0.02 \\
        Kepler-539	&5767	&12	&4.49	&0.03	&1.47	&0.03 & -0.10 & 0.01 \\
    \enddata
\end{deluxetable}

\subsection{Stellar Abundances \label{sec:stellar_abuns}}
Abundances of C, O, Mg, Si, and Ni in our sample of stars were determined by combining measurements of lines corresponding to these elements (and in the case of C, CH and C$_{2}$ molecular features) and the best-fit stellar parameters derived above in a curve-of-growth analysis within MOOG using the ``abfind'' package. The line lists for these elements are drawn from \cite{ramirez2014} and \cite{teske2014}. The line EWs measured in the each host star and its corresponding solar reference -- using the same groupings as in the Fe I and II line measurements -- are listed in Table \ref{tab:lines}. As with the iron lines, our procedure for measuring the EWs in each spectrum consisted of a line-by-line comparison between the solar reference spectrum and the normalized and Doppler-corrected spectrum of the star of interest, fitting a Gaussian profile to each line with the \texttt{splot} task in IRAF. This allowed us to choose a continuum region that was the roughly same in the host star and solar spectra, reducing potential systematic error. For carbon and oxygen, we also applied a synthesis analysis -- comparing lines and a small surrounding continuum region in the observed spectrum to the same region in synthesized stellar atmospheric models with the appropriate stellar parameters -- to two molecular C$_{2}$ lines at 5086 and 5135~{\AA} and the forbidden [O I] line at 6300~{\AA}. The details of the carbon and oxygen abundance determination are given below in Sections \ref{sec:carbon} and \ref{sec:oxygen}. 

For the five of stars that we analyzed completely by-hand -- WASP-8, HAT-P-15, HAT-P-17, Kepler-89, and HD 80606 -- we used the same prescription for the abundance error determination as in \cite{teske2014}. These uncertainties incorporate the change in the elemental abundances caused by perturbing the adopted stellar parameters ($T_{\rm{eff}}$, log $g$, and $\xi$) within their respective uncertainties. The sensitivity of the abundance to each parameter was calculated for changes of $\pm$150~K in $T_{\rm{eff}}$, $\pm0.25$ dex in log $g$, and $\pm0.30$ km s$^{-1}$ in $\xi$, and then scaled by the actual uncertainty in each parameter for a given star. The dispersion in the abundances derived from different lines is parameterized with the uncertainty in the mean, $\sigma_{\mu}$\footnote{$\sigma_{\mu}$ = $\sigma / \sqrt{(N-1)}$, where $\sigma$ is the standard deviation and $N$ is the number of lines measured.}. Then the total uncertainty for each abundance is the quadratic sum of this dispersion and the three scaled parameter uncertainties. A similar abundance error prescription is used for the rest of the stars that we analyzed with the more automated $q^2$, except that the stellar parameter contribution is calculated by computing the changes in abundance caused by the exact errors in $T_{\rm{eff}}$, log $g$, and $\xi$, rather than by scaling sensitivities. 

The relative elemental abundances, averaged over all of the lines (or groups of lines, in the case of C and O), and their respective errors for each host star are listed in Tables \ref{tab:abuns_mgsifeni} and Table \ref{tab:abuns_co}.

\begin{deluxetable}{lcccccccccc}
    \tablecolumns{11}
    \tablewidth{0pc}
    \tabletypesize{\scriptsize}
    \tablecaption{Stellar Abundances of Fe, Mg, Si, and Ni \label{tab:abuns_mgsifeni}}
    \tablehead{
        \colhead{Star} & \colhead{[Fe I/H]} & \colhead{[Fe I/H] error} & \colhead{[Fe II/H]} & \colhead{[Fe II/H] error} & \colhead{[Mg/H]} & \colhead{[Mg/H] error} & \colhead{[Si/H]} & \colhead{[Si/H] error} & \colhead{[Ni/H]} & \colhead{[Ni/H] error}}
    \startdata
        WASP-8 & 0.286 & 0.059 & 0.293 & 0.045 & 0.250 & 0.033 & 0.273 & 0.015 & 0.315 & 0.033 \\		
        HAT-P-15 & 0.272 & 0.031 & 0.276 & 0.053 & 0.212 & 0.060 & 0.265 & 0.012 & 0.302 & 0.018 \\ 	
        HD80606	& 0.274 & 0.051 & 0.275 & 0.122 & 0.328 & 0.031 & 0.328 & 0.017 & 0.300 & 0.030 \\
        HAT-P-17 & 0.021 & 0.033 & 0.019& 0.094 & 0.062 & 0.045 & 0.060 & 0.025 & 0.034 & 0.019 \\	
        Kepler-89& 0.050 & 0.089 & 0.020 & 0.069 & -0.002 & 0.054 & 0.071 & 0.026 & 0.039 & 0.052 \\	
        Kepler-419& 0.032 & 0.141 & 0.034 & 0.079& 0.010 & 0.051 & 0.117 & 0.026 & 0.021 & 0.047 \\	
        Kepler-432& 0.016 & 0.063 & 0.009 & 0.078 & 0.005 & 0.046 & 0.070 & 0.022 & 0.027 & 0.025 \\
        WASP-84 &   0.048 & 0.069 & 0.054 & 0.083 & 0.014 & 0.015 & 0.066 & 0.016 & 0.038 & 0.018 \\ 		
        WASP-139&  	0.088 & 0.045 & 0.091 & 0.111 & 0.152 & 0.030 & 0.073 & 0.018 & 0.102 & 0.014 \\
        K2-24	& 0.371 & 0.037 & 0.367 & 0.049 & 0.437 & 0.045 & 0.420 & 0.015 & 0.458 & 0.015 \\   
        WASP-130& 0.285 & 0.037 & 0.284 & 0.035 & 0.269 & 0.029 & 0.322 & 0.010 & 0.340 & 0.011\\	
        WASP-29	&0.118 & 0.089 & 0.117 & 0.342 & 0.146 & 0.026 & 0.145 & 0.041 & 0.140 & 0.037 \\	
        Kepler-9 & 0.055 & 0.030 & 0.57 & 0.025 & -0.026 & 0.046 & 0.041 & 0.008 & 0.018 & 0.010\\	
        K2-19& 0.034 & 0.045 & 0.028 & 0.059 & -0.032 & 0.048 & 0.037 & 0.012 & 0.015 & 0.012 \\		
        K2-27& 0.111 & 0.048 & 0.116 & 0.084 & 0.106 & 0.022 & 0.123 & 0.014 & 0.136 & 0.014\\		
        K2-139 (EPIC 218916923)	& 0.188 & 0.047 & 0.193 & 0.112 & 0.148 & 0.020 & 0.213 & 0.013 & 0.188 & 0.014\\
        Kepler-145& -0.034 & 0.039 & -0.033 & 0.039 & -0.102 & 0.016 & 0.037 & 0.012 & -0.022 & 0.018 \\
        Kepler-277& 0.042 & 0.045 & 0.039 & 0.045 & 0.057 & 0.038 & 0.085 & 0.048 & 0.062 & 0.021 \\	
        CoRoT-9 & -0.014 & 0.031 & -0.015 & 0.034 & 0.007 & 0.006 & -0.028 & 0.035 & -0.036 & 0.009 \\		
        Kepler-539& -0.096 & 0.028 & -0.094 & 0.030 & -0.040 & 0.009 & -0.099 & 0.032 & -0.164 & 0.010\\
    \enddata
\end{deluxetable}

\begin{splitdeluxetable*}{lcccccBlcccccccc}
    \tablecolumns{15}
    \tablewidth{0pc}
    \tabletypesize{\scriptsize}
    \tablecaption{Stellar Abundances of C and O \label{tab:abuns_co}}
    \tablehead{
        \colhead{Star} & \colhead{[C/H]}  & \colhead{[C/H] error} & \colhead{[C/H]$_{\rm{C2}}$} & \colhead{[C/H] avg} & \colhead{[C/H] spread} & \colhead{Star} & \colhead{[O/H]$_\mathrm{forb}$} & \colhead{[O/H]$_\mathrm{forb}$ error} & \colhead{[O/H]$_\mathrm{forb}$ synth} & \colhead{[O/H]$_\mathrm{triplet}^{c}$} & \colhead{[O/H]$_\mathrm{triplet}$ error} & \colhead{[O/H] avg} & \colhead{[O/H] quad-sum error} & \colhead{[O/H] spread}}
    \startdata
        WASP-8 & 0.123 & 0.057 & 0.175 & 0.149 & 0.052 & WASP-8 & 0.155 & 0.054 & 0.150 & 0.154 & 0.066 & 0.153 & 0.060 & 0.001 \\
        HAT-P-15 & 0.155 & 0.066  & 0.210 & 0.182 & 0.056 & HAT-P-15 & 0.100 & 0.035 & 0.110 & 0.145 & 0.050 & 0.118 & 0.043 & 0.045 \\
        HD80606 & 0.270 & 0.098 &    &       &       & HD80606 & 0.200 & 0.083 &       &        &       &      &        &       \\
        HAT-P-17 & 0.075 & 0.069  & 0.030 & 0.053 & 0.045 &HAT-P-17& -0.007 & 0.084 & 0.020 & 0.045 & 0.061 & 0.019 & 0.073 & 0.051 \\
        Kepler-89 & -0.054 & 0.164 & 0.050 & -0.002 & 0.104 &Kepler-89&   &  &       &  & &        &    &    \\
        Kepler-419 & 0.131 & 0.100$^{a}$ &      &   &             &Kepler-419    &         &       &  &0.131 & 0.108&     &       &    \\
        Kepler-432 & -0.103 & 0.084$^{a}$ &     &   &               &Kepler-432& -0.104 & 0.045 & -0.11 & -0.040 & 0.072& -0.085 & 0.060 & 0.070 \\
        WASP-84 & 0.131 & 0.043  & -0.060 & 0.036 & 0.191 & WASP-84& -0.147 & 0.053 & -0.050 & 0.077 & 0.049 & -0.03 & 0.051 & 0.224 \\
        WASP-139 &  &  0.15$^{b}$ & 0.130 & & & WASP-139&-0.011 & 0.028 & 0.030 & 0.025 & 0.027 & 0.015 & 0.027 & 0.036 \\
        K2-24  &  0.220 & 0.063 & 0.360 & 0.290 & 0.140 & K2-24&0.284 & 0.028 & 0.360 &0.286 &0.065 &0.310 &0.050 &0.002 \\
        WASP-130& 0.237 & 0.030 & 0.310 & 0.274 & 0.073 &WASP-130& 0.086& 0.020&0.220 &0.231 &0.069 &0.179 &0.051 &0.145 \\
        WASP-29 &       &   0.15$^{b}$  & 0.140 &  &  & WASP-29 & 0.028&0.038&0.070 &0.024 &0.106 &0.041 &0.080 &0.004 \\
        Kepler-9 & -0.061 & 0.026 & -0.065 & -0.063 & 0.004&Kepler-9 & 0.057& 0.017&0.080 &0.066 &0.016 &0.068 &0.017 &0.009 \\
        K2-19    & -0.043 & 0.041 &-0.095 & -0.069 & 0.052& K2-19 & -0.070&0.026 &-0.040 &0.006 &0.035 &-0.035 &0.031 &0.076 \\
        K2-27    & 0.103 & 0.034 & 0.065 & 0.084 & 0.038& K2-27 & & & & 0.042&0.047$^{a}$& & & \\
        K2-139 (EPIC 218916923) & 0.258 & 0.060 &0.165 & 0.212 & 0.093 &K2-139&   &   &  &0.149 &0.047$^{a}$ & & & \\
        Kepler-145 & -0.029 & 0.024 &0.020 & -0.010 &  0.110&Kepler-145 & -0.029& 0.024& 0.020 &-0.009&0.011 & -0.005 & 0.019 & 0.029 \\
        Kepler-277 &0.082 & 0.049  &0.000  & 0.041  &  0.082 & Kepler-277& & & & 0.086& 0.041& & & \\
        CoRoT-9   &-0.037 & 0.016  &-0.049  &-0.111 &  0.001& CoRoT-9 & -0.037&0.016 &-0.040 & -0.045&0.020 &-0.041 &0.018 &0.008 \\
        Kepler-539& -0.156 & 0.021  &-0.220 &-0.188 &  0.064& Kepler-539 & 0.024&0.015 &-0.090 &-0.074 &0.022 &-0.047 &0.019 &0.114 \\
    \enddata
    \tablenotetext{a}{Errors increased to include potential systematic error induced by measuring a limited number of abundance indicators. See the appendix for further details.}
    \tablenotetext{b}{Errors conservatively estimated from [C/Fe] vs. [Fe/H] relation in the study of \cite{nissen2014} of solar twins.}
    \tablenotetext{c}{All values reflect NLTE corrections.}
\end{splitdeluxetable*}

\subsubsection{Carbon Abundance Derivation \label{sec:carbon}}
We used in our abundance derivation four C I lines and two molecular C$_2$ features at 5086 and 5135~{\AA} (see Table \ref{tab:lines}). We derived [C/H] from the C I lines using EW measurements, striving for a strictly differential analysis by keeping the continuum regions consistent between the solar reference and host star spectra. The four C I lines -- 5052.17, 5380.34, 6587.61, and 7111.48~{\AA} -- were selected based on the wavelength coverage of our spectra and on the high-quality factor of the lines, as described by Caffau et al. \citeyearpar[][see their Tables 1 and 3]{caffau2010}. However, the actual quality of each line varies in the spectra of different stars (depending on stellar parameters and/or S/N), so not every star has measurements for all four C I lines. For HD 80606, which was reported in \cite{teske2014}, we did not make new EW measurements of the C I lines, but simply excluded the 7113.18~{\AA} line and recomputed the average and spread of [C/H].

We also derived [C/H] from two blended components of the C$_2$ Swan system, at 5086 and 5135~{\AA}, via a spectral synthesis analysis (matching a synthetic spectra to the observed spectrum). The line lists used to create the synthetic spectra are from \cite{schuler2011}, with C$_2$ molecular data from \cite{lambert&ries1981} and the dissociation energy (6.297 eV) from \cite{urdahl1991}. Each synthesized spectrum was convolved with a Gaussian function, based on nearby unblended lines, to approximate the point-spread function (PSF) of the instrument and any additional broadening not accounted for by the stellar parameters described above. The free parameters were then the relative continuum flux, wavelength shift, and carbon abundance, and the best fit was determined by minimizing the differences between the observed and synthetic spectra. As was the case for the C I atomic lines, not every star has measurements for the C$_{2}$ lines, depending on the quality of their respective spectra in those regions.

The errors on [C/H] derived from the atomic lines were determined by the procedure described above for the rest of the elements. We did not derive a formal error on the [C/H] values derived from the C$_2$ synthesis fits. Instead, we report the spread between the [C/H] values derived from the atomic versus from the molecular lines, and use as our final [C/H] error the higher of the two values between the spread and the formal [C I/H] error. Our final [C/H] value is the average between the atomic and molecular abundance determinations. 

In two cases -- WASP-139 and WASP-29 -- only one C I line (5380.34~{\AA}) was readily measurable in the stellar spectra. The resulting [C I/H] values derived from this line were very high, 0.263$\pm$0.021 and 0.686$\pm$0.073 for WASP-139 and WASP-29, respectively, given the stellar [Fe/H] values (0.09$\pm$0.02 and 0.12$\pm$0.05 for WASP-139 and WASP-29, respectively). Thus, we chose to use only the C$_2$ features as carbon abundance indicators in our analysis of these stars. Because we did not measure formal errors for the molecular features, we instead estimated conservative [C/H] uncertainties ($\pm$0.15 dex for both stars) of WASP-139 and WASP-29 based on the stellar [Fe/H] values and the spread of [C/Fe] values in the study of solar twins by Nissen et al. \citeyearpar[][see their Figure 9]{nissen2014}.

The two most evolved stars in our sample, Kepler-419 and Kepler-432, have C$_2$ features that are too broad to be reliable carbon abundance indicators. In these cases we only used the C I lines in our abundance determination, specifically, only the two bluer C I lines. These lines are of the highest quality \citep{caffau2010}, and the two redder C I lines result in [C/H] abundances that are higher by $\sim$0.1-0.25 dex versus the average of the two bluest C I lines. 

\subsubsection{Oxygen Abundance Derivation \label{sec:oxygen}}
To derive oxygen abundances for our sample of stars, we used two main indicators -- the [O I] forbidden line at 6300~{\AA}, and the O I triplet at $\sim$7775~{\AA}. The forbidden line is weak, but does not suffer non-local thermal equilibrium (non-LTE) effects like the triplet lines \citep[e.g.,][]{kiselman1991,kiselman1993,gratton1999}. However, the [O I] line is blended with two isotopic components of a Ni I transition, which represent $\sim$55\% of the strength of the [O I] line in the Sun \citep{caffau2008}.\footnote{The [O I] line can also suffer telluric contamination depending on the velocity of the star relative to Earth. We did not account for any such contamination in our analysis. However, as described in Appendix \ref{sec:offsets}, the errors on our final oxygen abundance measurements account for differences between the forbidden and the NTLE-corrected triplet lines. These errors should encompass the  uncertainty introduced by telluric contamination in the forbidden line.} To account for this blend, we first use the \texttt{blends} driver in MOOG, which takes as input the EW, log $gf$ (-9.717, \citealt{storey&zeippen2000}), and excitation potential ($\chi$=0) of the [O I] line, as well as the atomic parameters for the Ni transitions, log $gf$($^{60}$Ni) = -2.965 and log $gf$($^{58}$Ni) = -2.275 and $\chi$=4.27 \citep{bensby2004}. We modified the best-fit stellar atmospheric models for each star to reflect our measured [Ni/H] value based on our EW analysis of $\sim$ 30 other Ni I lines. 

As a check on the \texttt{blends} driver, we also performed a synthesis analysis on the [O I] line, similar to the analysis of the C$_2$ molecular features above. The line list used in the synthesis analysis is the same as in \cite{teske2014}, provided via private communicate by V. V. Smith, and includes CO, TiO, as well as several atomic features spanning $\sim$6295-6305~{\AA}. The synthesis line list includes the same line parameters for the forbidden oxygen line and the Ni isotopic components as in the \texttt{blends} driver analysis. Similar to the C$_2$ synthesis analysis, we do not derive a formal error on the oxygen abundance measured via this synthesis analysis.

The O I triplet is composed of the unblended relatively strong lines at 7771.94, 7774.17, and 7775.39~{\AA}, but they suffer from NLTE effects (see description in \citealt{teske2013} and references therein). Under the assumption of local thermal equilibrium (LTE), the abundances derived from these lines are overestimated. There exist multiple correction methods in the literature that establish the magnitude of departure from LTE, usually parameterized by relations involving the stellar parameters \citep[e.g.][]{takeda2003, ramirez2007,fabbian2009}. In this work we mostly use \cite{ramirez2007}, as it is built into the $q^2$ code. For four of the five stars that we analyzed by hand as described above (WASP-8, HAT-P-15, HAT-P-17, and Kepler-89), we calculated and applied NLTE corrections from \cite{takeda2003}, \cite{ramirez2007}, and \cite{fabbian2009}, and took the resulting average value as the final [O/H]$_{\rm{triplet}}$ value; the differences between the averages and the \cite{ramirez2007}-only [O/H]$_{\rm{triplet}}$ values are within the quoted errors. For HD 80606, we simply use the [O/H] abundance reported in \cite{teske2014}, which was only derived from the [O I] line. 

In our analysis below, we use as a conservative error on [O/H] the higher of two values: the spread across the abundance indicators described above (``[O/H] spread'' in Table \ref{tab:abuns_co}), and the quadrature sum error $\sqrt{err_\mathrm{forb}^2 + err_\mathrm{triplet}^2/2}$ (``[O/H] quad-sum error'' in Table \ref{tab:abuns_co}). In a few cases (Kepler-419, K2-27, K2-139, and Kepler-277) we were unable to resolve the [O I] line, therefore we rely only on the [O/H]$_{\rm{triplet}}$ values and errors. In the case of Kepler-89, the [O I] line was unresolved, and the EW measurements of the oxygen triplet lines resulted in a potentially unrealistically low [O/H] value ($\sim$-0.40 dex) given the [Fe/H] of the star (0.05$\pm$0.11 dex). Thus, for Kepler-89 we do not report an [O/H] value; perhaps higher S/N data would make this measurement more feasible or plausible. 

Interestingly, \cite{kiselman1993} suggest granulation as one factor in the discrepancy between oxygen abundances derived from the oxygen triplet assuming LTE and other line indicators. Granulation velocities are expected to be more vigorous in earlier type/more luminous stars like Kepler-89 (late-F spectral type). However, we note that the difference between the [Fe/H] (0.032$\pm$0.51 dex) and [O/H]$_{\rm{triplet}}$ values (0.131$\pm$0.108) for Kepler-419 is not large. This star is also a hot, luminous star.

\begin{figure}[p]
    \centering
    \begin{minipage}{.47\textwidth}
    	\centering
    	\vspace{-15pt}
    	\includegraphics[width=\textwidth,clip]{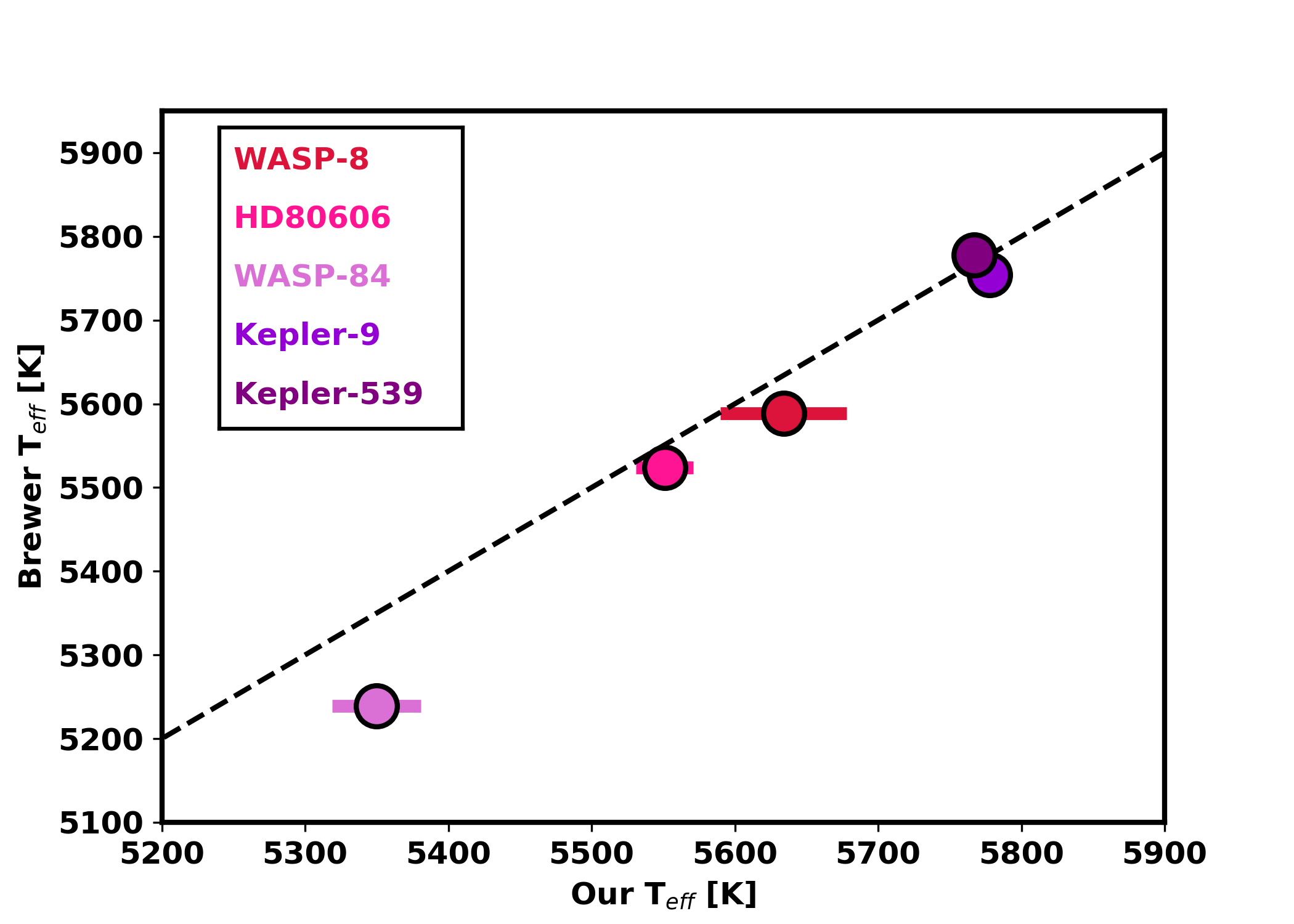} 
    \end{minipage}
    \begin{minipage}{0.47\textwidth}
    	\centering
    	\vspace{-15pt}
    	\includegraphics[width=\textwidth,clip]{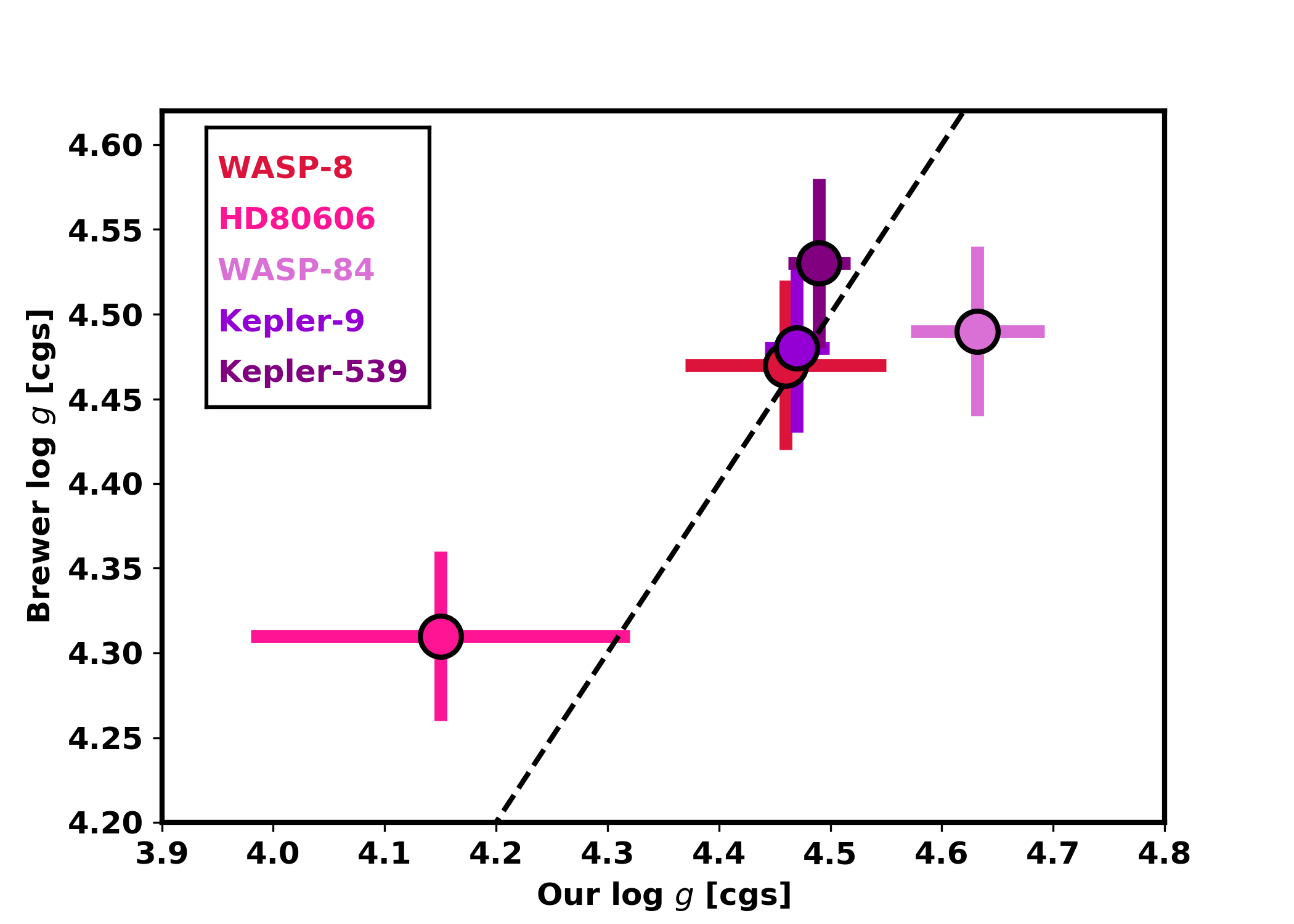} 
    \end{minipage}
    \begin{minipage}{0.47\textwidth}
    	\centering
    	\vspace{-15pt}
    	\includegraphics[width=\textwidth,clip]{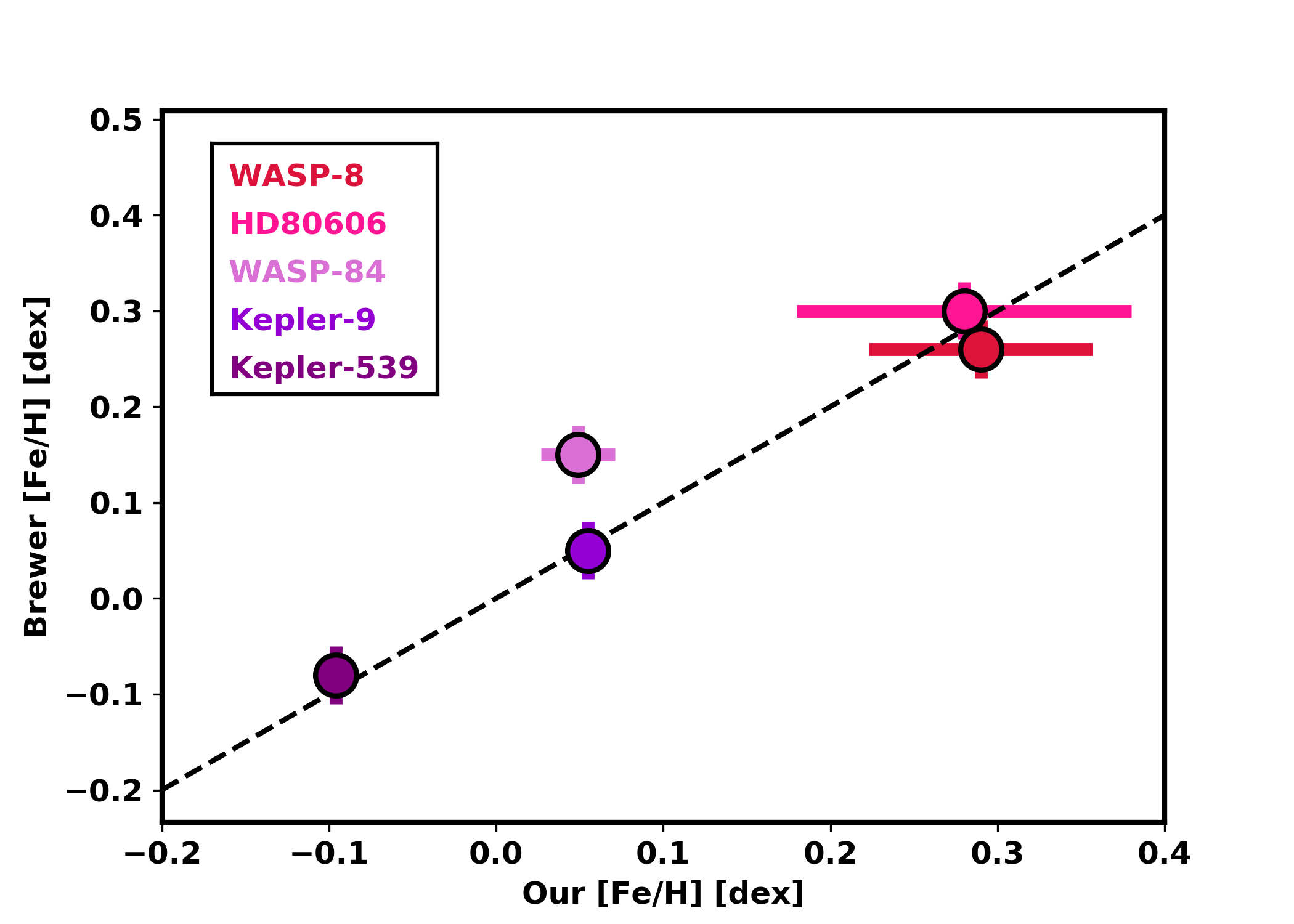} 
    \end{minipage}
    \begin{minipage}{.47\textwidth}
    	\centering
    	\vspace{-15pt}
    	\includegraphics[width=\textwidth,clip]{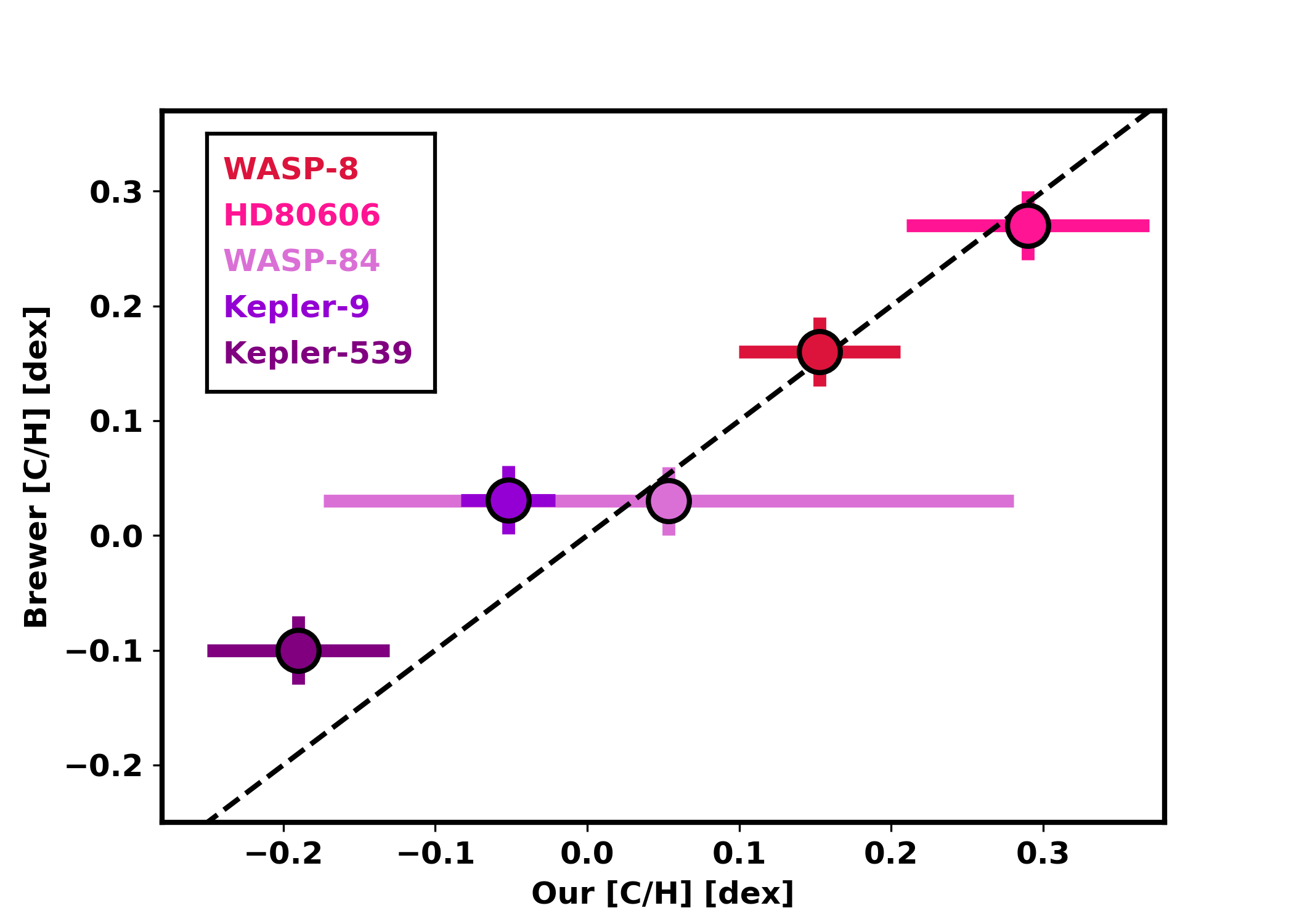} 
    \end{minipage}
    \begin{minipage}{0.47\textwidth}
    	\centering
        \vspace{-15pt}
    	\includegraphics[width=\textwidth,clip]{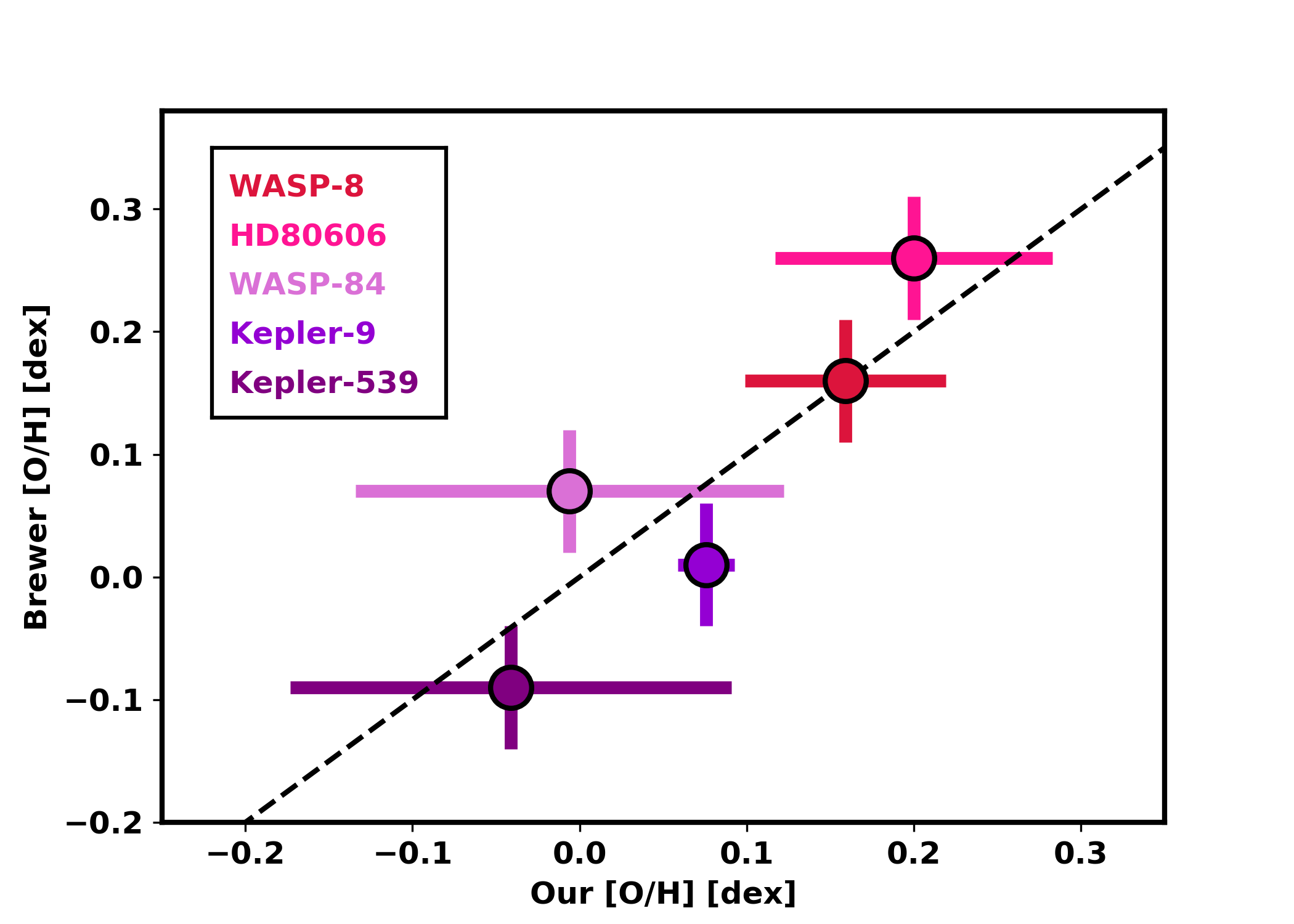} 
    \end{minipage}
    \begin{minipage}{0.47\textwidth}
    	\centering
        \vspace{-15pt}
    	\includegraphics[width=\textwidth,clip]{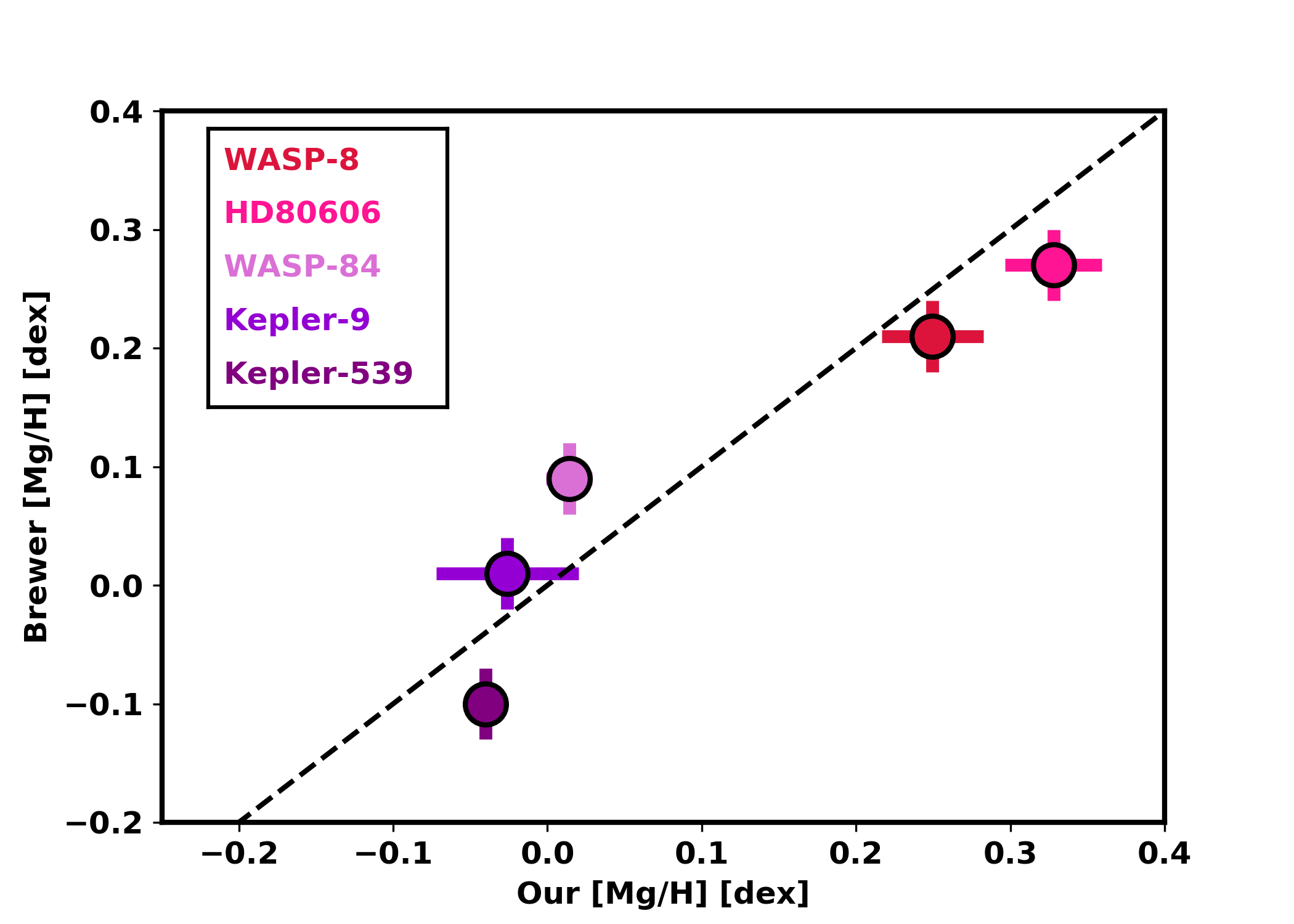} 
    \end{minipage}
    \begin{minipage}{0.47\textwidth}
    	\centering
        \vspace{-15pt}
    	\includegraphics[width=\textwidth,clip]{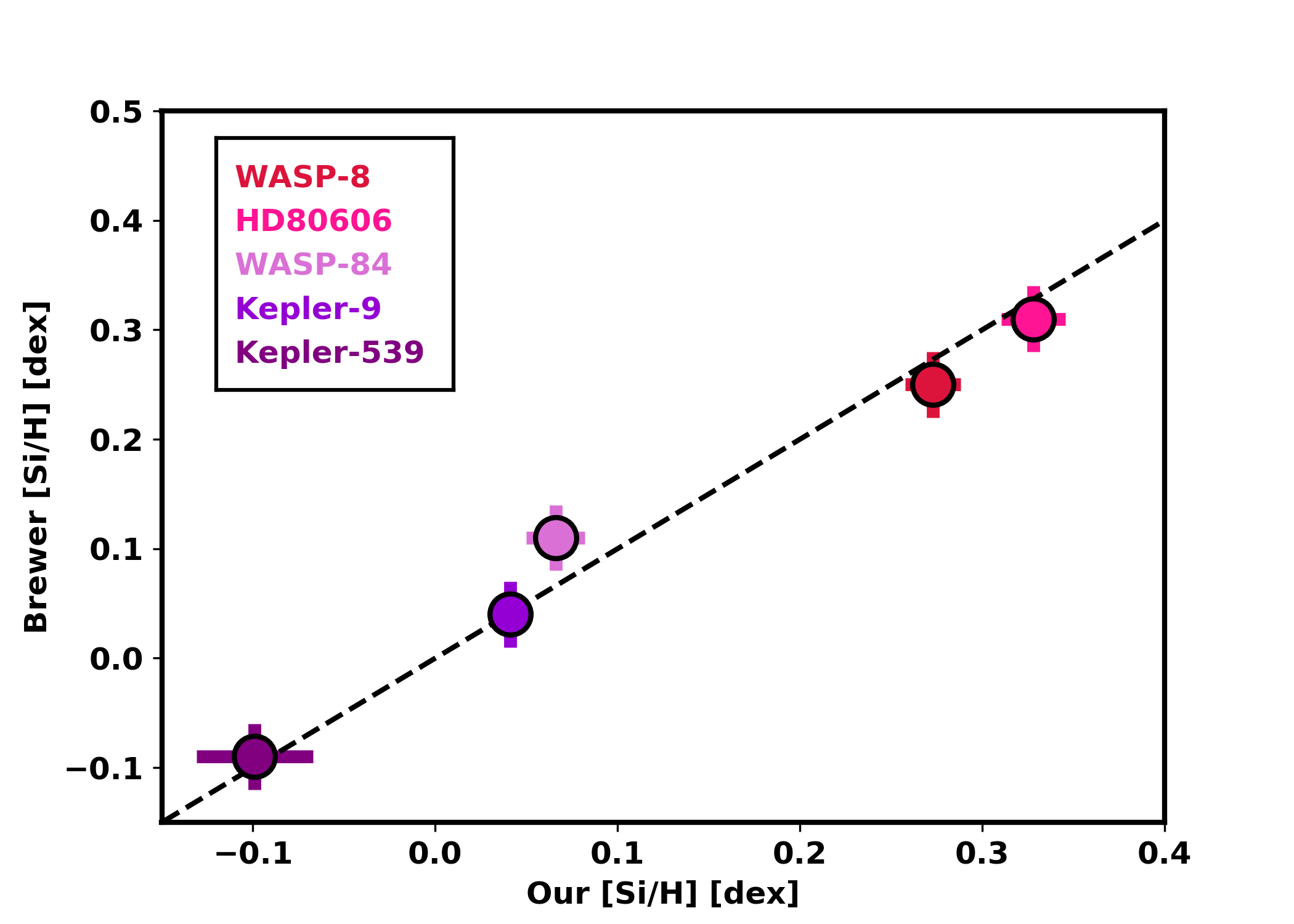} 
    \end{minipage}
    \begin{minipage}{0.47\textwidth}
    	\centering
        \vspace{-15pt}
    	\includegraphics[width=\textwidth,clip]{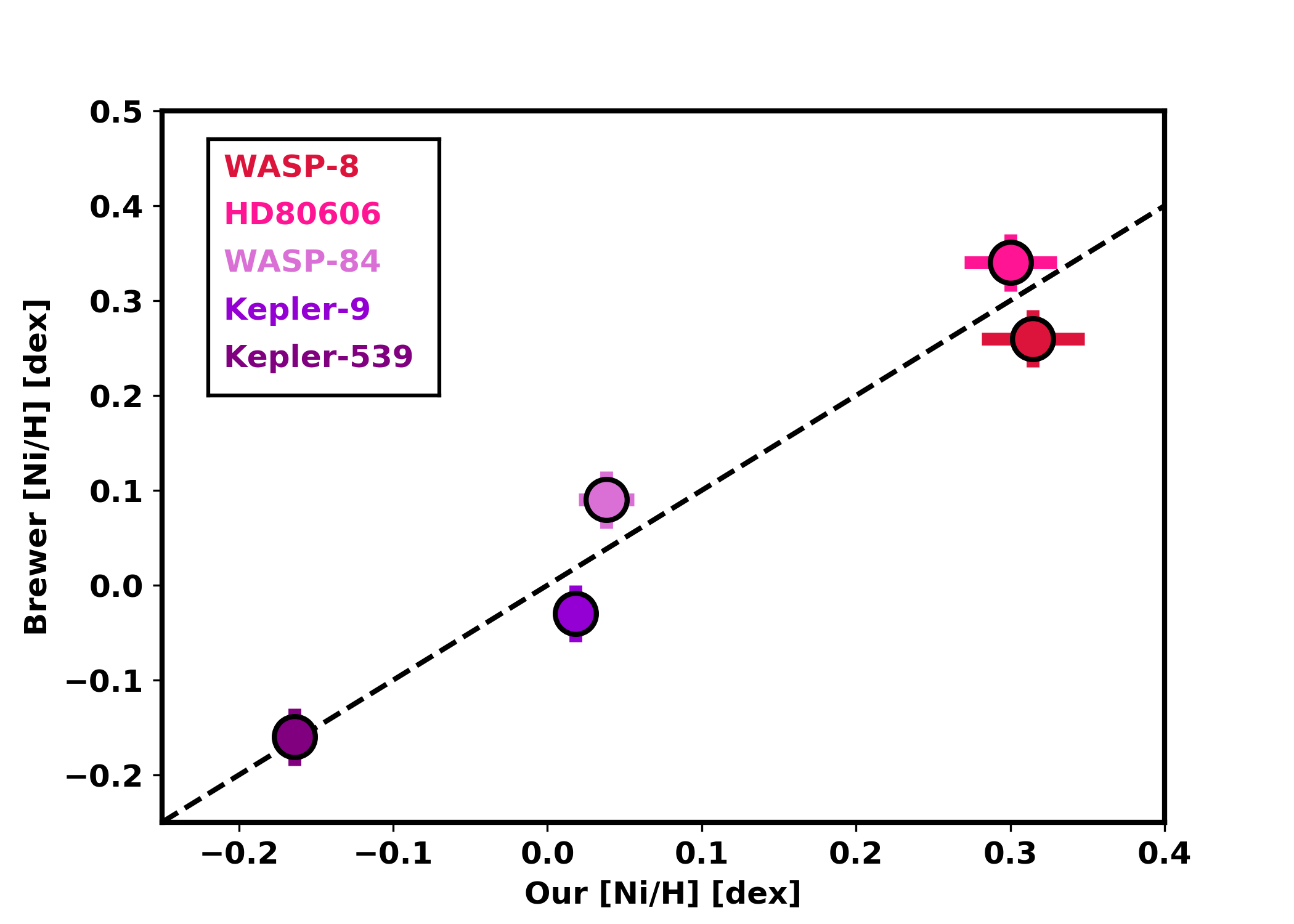} 
    \end{minipage}
    \caption{Comparison of stellar parameters and abundances derived in this work versus those derived by \cite{brewer2018} for five host stars in our sample. The dashed lines represent the 1:1 correlation. There is good agreement between the two studies, giving us confidence in incorporating measurements of four additional stars from \cite{brewer2018} into our analysis.}
    \label{fig:brewer comparison}
\end{figure}

The inhomogeneous derivation of the [C/H] and [O/H] abundances -- not all lines are measured in all stars -- could introduce systematic bias into our abundances, thus impacting our measured correlations. To attempt to account for bias, we applied an empirically derived correction to the stars with missing abundance indicators for carbon and oxygen, as detailed in the appendix. Including these corrections does not significantly change our results, but may still not fully account for systematic bias. Thus we also performed a sensitivity test, repeating our analysis after excluding from the sample the stars that lacked the most frequently measured C and O indicators (C I and the oxygen triplet) and Kepler-282 and Kepler-238 (see \S\ref{sec:BF18_compare}). The details of the sensitivity analysis are also described in the appendix, and based on it, we also found that our final conclusions did not change.

\subsection{Comparison to the Results of Brewer \& Fischer (2018)  \label{sec:BF18_compare}}
The recent results from \citeauthor{brewer2018} (2018, \citetalias{brewer2018}), who produced a catalog of 15 elemental abundances for stars observed by the California-\textit{Kepler} Survey, provide an opportunity for a consistency check on the abundances derived here. Briefly, \citetalias{brewer2018} analyzed Keck I/HIRES spectra of similar resolution and wavelength range as those analyzed here, using Spectroscopy Made Easy (SME, \citealt{Piskunov&Valenti2017}) to first fit a single synthetic stellar model across the optical spectrum and derive global stellar parameters and the $\alpha$-elements Ca, Si, and Ti. The authors then perturbed the effective temperature of the stellar model by $\pm$100 K, refit the spectra, and fixed the stellar parameters to the average values from all three fits. Then the authors fit for 15 elements, including C, O, Mg, Si, Fe, and Ni, repeated the entire procedure once more, and used the results of the second iteration as the final reported values. 

In Figure \ref{fig:brewer comparison} we show a comparison of their stellar parameters ($T_\mathrm{eff}$, log~$g$, and [Fe/H]) and abundances ([C/H], [O/H], [Mg/H], [Si/H], and [Ni/H]) to those derived in this work for five overlapping stars. In these plots, the dashed line represents the one-to-one correlation. There is good agreement between the stellar parameters and abundances. In a few instances no overlap occurs when the errors are considered, mostly for WASP-84, for which \citetalias{brewer2018} find slightly higher abundances for all elements except [C/H] (and the [C/H] value for WASP-84 derived in this work has a large error). However, the differences for [Fe/H], [Mg/H], [Si/H], and [Ni/H] are all $\leq$0.05 dex when the respective errors are considered, which is realistic for two studies using different techniques (synthetic spectra fitting versus EW measurement) and spectra (Keck I/HIRES for \citetalias{brewer2018} and \textit{Magellan} II/MIKE for WASP-84 in this study). For example, in a study of stellar abundances measured for the same stars across 84 literature sources \citep{Hinkel2017}, the spread in [Si/H] across 10 different papers using different methodologies and line lists was $\sim$0.25 dex (see their Figure 3). Indeed, the overall small differences between parameters and abundances of \citetalias{brewer2018} and those derived here are better than average. 

This comparison gave us the confidence to include in our analysis below two 
additional systems from \citetalias{brewer2018} that fall within the parameter space of interest to this work -- host stars of planets with $T_\mathrm{eq}<$ 1000~K, incident flux $<2 \times 10^{8}$ erg~s$^{-1}$~cm$^{-2}$, and measured $M_p$ and $R_p$ with errors $\leq50$\%. These stars are  
Kepler-282 and Kepler-238, and their parameters and abundances can be found in \citetalias{brewer2018}. 

\section{Analysis} \label{sec:analysis}
The point of this study is to understand how the bulk metallicities of giant planets are connected to the metal abundances of their host stars. To do this, we conducted correlation tests and Bayesian linear fits on the data using the Markov chain Monte Carlo (MCMC) method. In this section, we describe these methods in detail and how we handle prior information and the existing known mass-metallicity trend from \cite{thorngren2016}.

Although our measurements of elemental abundances are independent of one another, abundances within a given star are well known to be strongly correlated.  This represents a strongly motivated and relevant prior that should be included in our analysis.  To do this, we used the Hypatia Catalog\footnote{www.hypatiacatalog.com} of stellar abundances \citep{Hinkel2014,Hinkel2017} to analyze nearby thin-disk planet-hosting stars with measured C, O, Mg, Si, Fe, and Ni abundances.  The distribution of these abundances was well represented by a multivariate normal distribution. Using this population prior, it is possible to compute a joint posterior distributions on the abundances in closed form:
\begin{equation}
	\mathbf{[Z/H]} \sim \mathcal{N}\left(
        \left((\bm{1\sigma_o})^{-1}\bm{\mu_o}+\bm{\Sigma_p}^{-1}\bm{\mu_p}\right)
        \left((\bm{1\sigma_o})^{-1}+\bm{\Sigma_p}^{-1}\right)^{-1},
        \left((\bm{1\sigma_o})^{-1}+\bm{\Sigma_p}^{-1}\right)^{-1}\,\,\,.
    \right)
\end{equation}
Here, $\bm{1}$ is the identity matrix, $\bm{\mu_o}$ and $\bm{\sigma_o}$ are the mean and variance vectors of the the observed abundances, and $\bm{\mu_p}$ and $\bm{\Sigma_p}$ are the mean and vector covariance materix of the population priors. This approach does not affect the individual abundances very much, but it does introduce appropriate covariances between them.  This is primarily useful for correctly handling combinations of abundances, such as the total stellar metal abundance [Z/H] or the C/O ratio.  For a linear weighed combinations of abundances (e.g. volatiles - refractories) of weights $\bm{W}$, the resulting distribution is the usual $\mathcal{N}(\bm{W} \cdot \bm{\mu},\bm{W \Sigma W^T})$.  For nonlinear combinations like [Z/H], uncertainties were determined numerically via sampling. We have tested our analysis without the Hypatia-based prior, and although its absence modestly increases the uncertainties for some linear combinations of abundances, it does not change our overall results.

We conducted a variety of linear fits investigating the relationship between the stellar and planetary metallicity.  We used the Bayesian approach to conduct the fits, in which a dependent variable $y_o$ with observational uncertainty $\sigma_y$ is explained by a slope $m$ times the regressor $x$ with observational uncertainty $\sigma_x$ plus a constant $b$.  An additional uncertainty residual to the fit, $\sigma_f$, is also included.  The likelihood is given by the resulting prediction error in $y$:
\begin{align}
	p(\bm{y}_o,x_o|m,b,\bm{x},\bm{\sigma_y},\bm{\sigma_x},\sigma_f) =
    	\prod_i \mathcal{N}\left(y_{o,i}\middle|m x_i +b,
        \sigma_{y,i}^2+\sigma_f^2)\right)\,\,.
\end{align}
For the fit parameters $m$ and $b$, we used flat uninformative priors (in the Bayesian notation, $p(m,b) \propto 1$) and the residual spread was $p\propto \sigma_f^{-2}$ -- these are the same priors implied by the frequentist least-squares approach.  The prior on $x$ was just the observed distribution, which has mean $x_o$ and uncertainty $\sigma_x$.  This gives the following posterior distribution:
\begin{align}
	p(\bm{m}, b, \bm{x},\sigma_f|\bm{y_o},\bm{\sigma}_x,\bm{\sigma_y}) &\propto
    	p(\bm{y}_o,x_o|m,b,\bm{x},\bm{\sigma_y},\bm{\sigma_x},\sigma_f)
    	p(m,b,\bm{x},\bm{\sigma_y},\bm{\sigma_x},\sigma_f)\\
    &\propto \sigma^{-2}
    	 \prod_i \mathcal{N}\left(y_{o,i}\middle|m x_i +b,
        \sigma_{y,i}^2+\sigma_f^2\right)
        \mathcal{N}\left(x_{o,i}\middle|x_i,\sigma_{x,i}\right)
\end{align}

This is the standard Bayesian approach to regression adapted to the case where there is uncertainty in both the independent and dependent variable.  In the limit where the $x$ uncertainties approach zero, we have the standard least-squares approach.  To take samples of this posterior, we used a Gibbs MCMC sampler \citep{Hastings1970,Geman1984}.  As a second approach to evaluating the relationships between variables, we also apply the Kendall's tau correlation test in each plot where we show a fit.  $\tau$ indicates the test statistic, and $p$ represents the two-sided p-value of the test rejecting the null hypothesis that there is no relationship between the variables.  $\tau$ varies between -1 (fully anti-correlated) and 1 (fully correlated), and 0 indicates no correlation. For this test, we neglect the uncertainties in $x$ and $y$.

\begin{figure}
    \includegraphics[width=\textwidth]{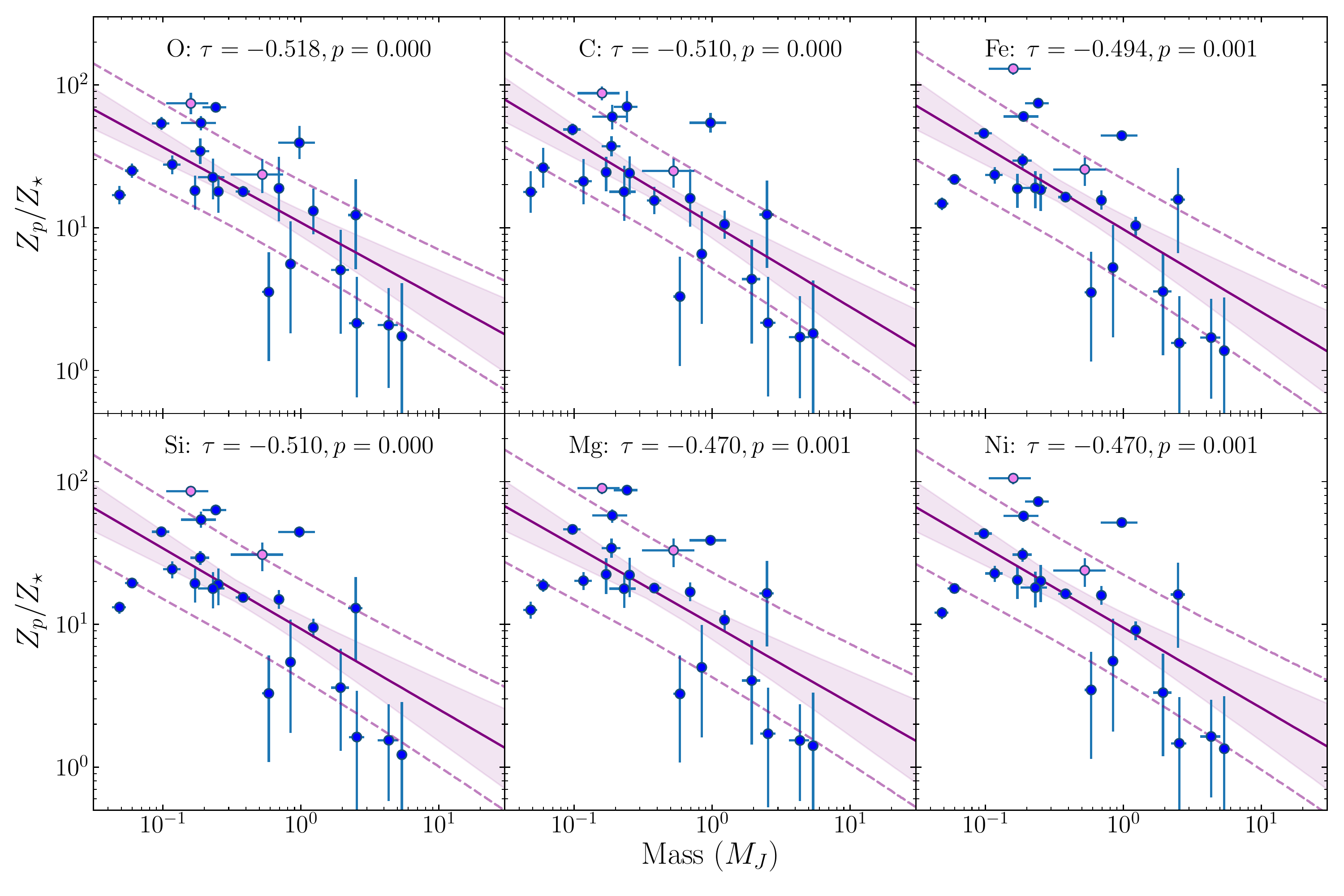}
    \caption{Scatter plots, fits, and Kendall's tau correlation tests depicting the relationship between the mass of a planet and its metal enrichment relative to the parent star using each element considered as a proxy for total the metallicity.  Shaded regions indicate the $1\sigma$ uncertainty in the fit, and dotted lines indicate the $1\sigma$ predictive interval. The abundances for the lighter points are from \cite{brewer2018}.  Our data produce the same negative relationship observed in \citetalias{thorngren2016} and \cite{miller&fortney2011}. }
    \label{fig:zzElement}
\end{figure}

We are primarily interested in correlations with planetary metallicity, which we estimate in the same way as in \citetalias{thorngren2016}: using thermal evolution models to match model planet radii to the observed radii by varying the bulk heavy-element content of the model planet.  Uncertainties in planetary metal content were again derived by varying the mass, radius, and age within their observed uncertainties and recording the resulting variance in planetary metal.  First, we consider the ratio of the planet to stellar metallicity by mass, $Z_p/Z_\star$, in line with previous analyses in \citetalias{thorngren2016} and \cite{miller&fortney2011}.  Figure \ref{fig:zzElement} shows the relationship between $Z_p/Z_\star$ and the planet mass using the various stellar metals as proxies for the total stellar metallicity. There is little difference in these trends across the elements explored in this paper.

In \citetalias{thorngren2016}, the authors estimated the heavy-element masses \textbf{$M_z$} for 47 planets and compared these values to the total planet mass $M$. The fit to this relationship, shown in their Figure 7, is $(57.9 \pm 7.03)~M^{(0.61 \pm 0.08)}$. Here we consider
the residual metal mass to that fit:
\begin{equation}
	\mathrm{Residual} = \frac{M_z}{58 \, M^{.61}}
\end{equation}
This quantity represents the relative amount of metal in a planet beyond (or below) what would be expected for its mass; on average, it should be $\sim$1 -- as it is for the entire \citetalias{thorngren2016} sample -- but there is some scatter around the fit from \citetalias{thorngren2016}, which results a scatter in residual metal mass. As such, how this variables relates to the stellar metal abundances is of particular interest to us.  Figure \ref{fig:residAbund} depicts the relationship between the measured stellar abundances and the residual metallicity for the planets. We discuss this figure in the next section in more detail. 

\section{Discussion} \label{sec:discussion}
\begin{figure}[b]
    \includegraphics[width=\textwidth]{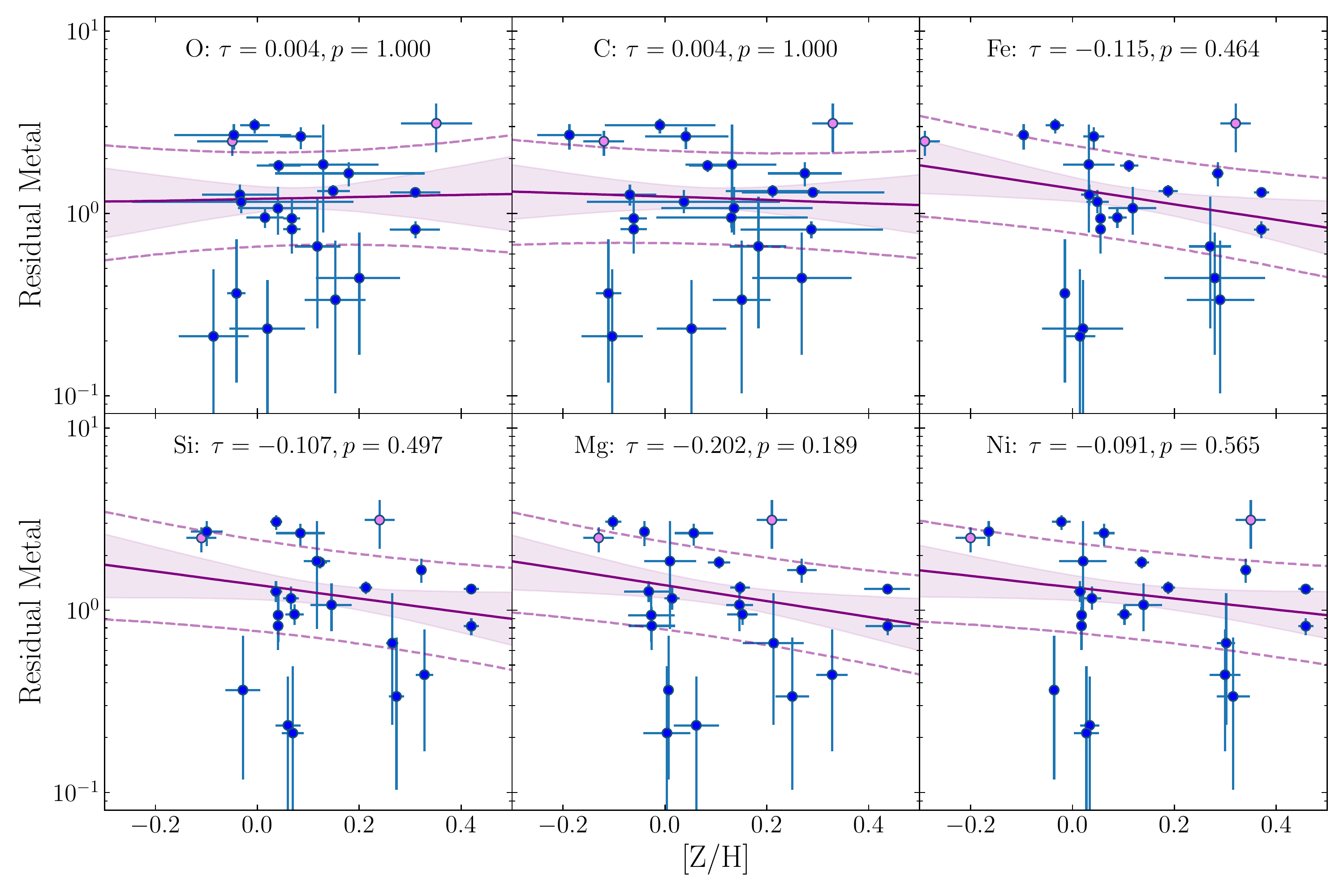}
    \caption{Scatter plots, fits, and Kendall's tau correlation tests depicting the relationship between the residual bulk metallicity of our planets and the stellar abundances of each of the considered elements.  Shaded regions indicate the $1\sigma$ uncertainty in the fit and dotted lines indicate $1\sigma$ predictive interval. The abundances for the lighter points are from \cite{brewer2018}.  Perhaps surprisingly, no significant relationship is apparent for any of the elements considered.}
    \label{fig:residAbund}
\end{figure}

\begin{figure}
    \includegraphics[width=\textwidth]{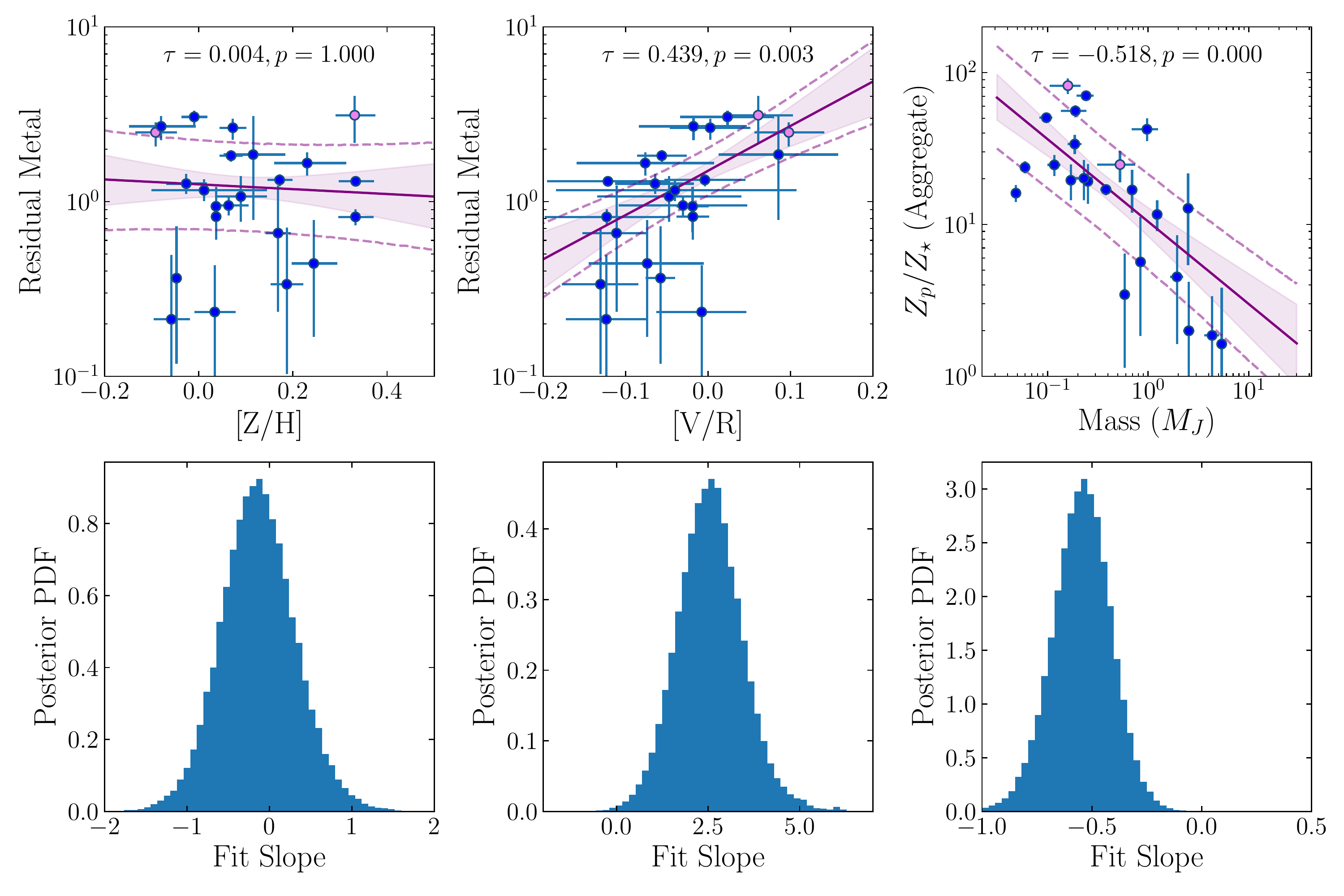}
    \caption{Plots, fits, Kendall's tau correlation tests, and slope histograms of the three major results of this paper.  Shaded regions indicate the $1\sigma$ uncertainty in the fit, and dotted lines indicate the $1\sigma$ predictive interval.  The abundances for the lighter points are from \cite{brewer2018}.  On the left, we show the lack of a strong correlation between the total metallicity of a star and the residual metallicity of the planet.  A slope of 1 is ruled out at the $2 \sigma$ level.  In the center, we show a correlation we observe between the residual metallicity of a planet (see \S \ref{sec:analysis}) and the ratio of stellar volatiles (C and O) to refractory elements (Fe, Si, Mg, and Ni).  On the right, we show that our data reproduce the relationship between planetary metal enrichment relative to the parent star and planet mass.  `Aggregate' indicates that we consider the total metallicity from all of the measured elements, rather than simply using one.}
    \label{fig:volRef}
\end{figure}

Strikingly, our results show an absence of a clear correlation between stellar and planetary (residual) metallicity (see Fig. \ref{fig:residAbund}).  Aggregating the metal abundances together (Fig. \ref{fig:volRef}, left panels), we can exclude a linear relationship (log-space slope of 1) between stellar and planetary abundances at the 2$\sigma$ level.  Such a linear relationship is a common prediction of formation models \citep[e.g.,][]{Mordasini2014a,thorngren2016}.  This represents the logic that if there is twice as much metal in the feeding zone of a forming giant planet, one would reasonably expect that twice as much metal would end up in the planet.  Additionally, it has been well established that the increased stellar metallicities are associated with a greater occurrence of giant planets \citep[e.g.,][]{fischer&valenti2005,Mordasini2008,Adibekyan2012}.

It is surprising that this positive relationship would not also be seen in the bulk metal abundances of these planets.  Perhaps more complex disk processes, such as the radial drift of grains \citep{Brauer2008,Birnstiel2012}, affect this relationship by regulating solid surface densities or creating metal-enhanced regions where the planets could form \citep{Powell2017,Yang2018}. Alternatively, an unmeasured but correlated parameter might be interfering with the trend.  For example, the orbit in which the planet formed in the disk (before migration) could be significant. Finally, our observed lack of a trend could even be evidence that the occurrence-metallicity association is not caused by planet formation per se \citep[e.g.,][]{Haywood2009}. These are only examples; the cause of this lack of linearity that we observe should be considered an open question in planet formation theory.

We observe an additional novel correlation with residual metal from our data.  From our separate measurements of six elemental abundances, we noted the comparison of volatile (C and O) vs. refractory elements in affecting planetary metal content. Because volatiles are relatively abundant \citep{Asplund2009} and can be found in both the gas and solid phases, we hypothesized that they may affect planetary metallicity differently.  In our data, we saw a correlation between residual metal and the ratio of volatile enrichment to refractory enrichment (Fig. \ref{fig:volRef}, middle panel, [$V/R$]).  

\begin{figure*}[ht!]
    \centering
    \begin{minipage}{.51\linewidth}
    	\centering
    	\vspace{-15pt}
    	\includegraphics[width=\textwidth,clip]{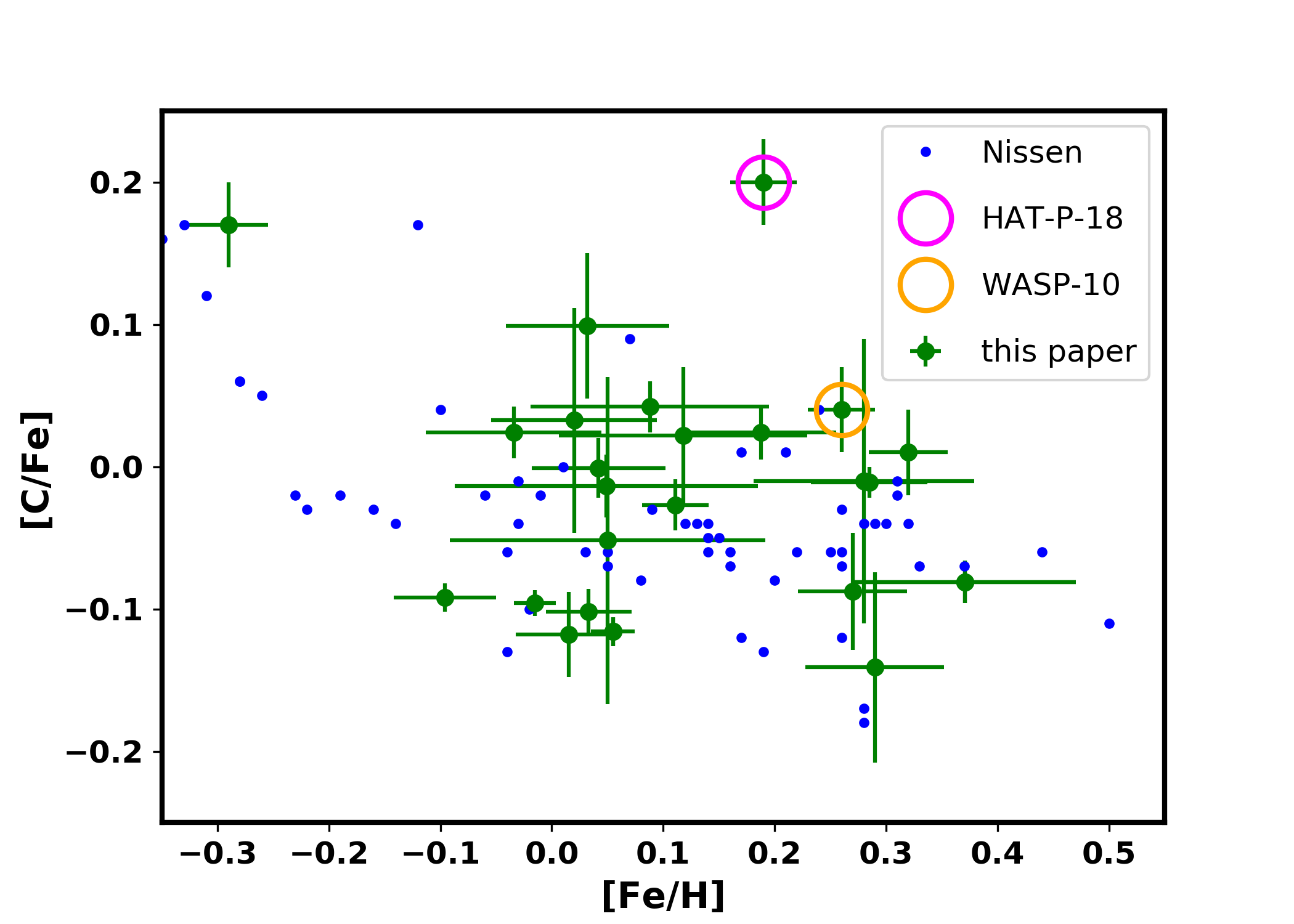} 
    \end{minipage}
    \hspace{-20pt}
    \begin{minipage}{0.51\linewidth}
    	\centering
    	\vspace{-15pt}
    	\includegraphics[width=\textwidth,clip]{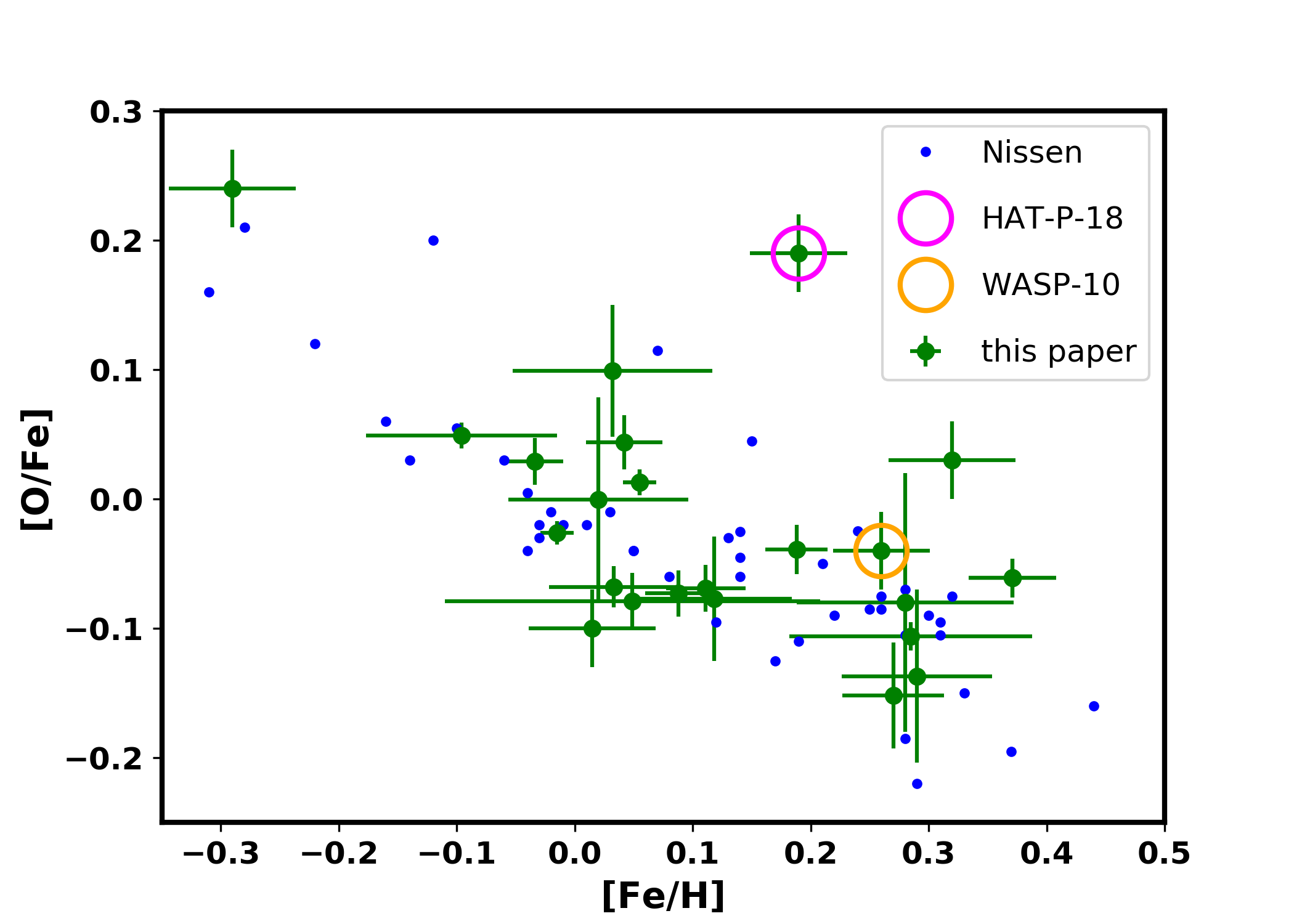}
    \end{minipage}
    \caption{Here we show trends in [C/Fe] (left) and [O/Fe] (right) with [Fe/H] that reflect the Galactic chemical evolution (GCE). The blue points are from the study of \cite{nissen2014} of solar twins, the green points are the data in this paper, and the larger magenta and orange circles highlight HAT-P-18 and WASP-10, two of the stars included in our study that were originally presented in \cite{brewer2018}. WASP-10 (orange circle) falls within the envelope of the GCE trends, but HAT-P-18 (magenta circle) appears to be an outlier, with high [C/Fe] and [O/Fe] for its [Fe/H].}
    \label{fig:gce}
\end{figure*}

This correlation should be interpreted with caution. There were two additional systems from \citetalias{brewer2018}  -- HAT-P-18 and WASP-10 -- that met our planet criteria described in \S\ref{sec:obs}. With the inclusion of these two planets, the trend in Figure \ref{fig:volRef} was eliminated by HAT-P-18 as a strong outlier. However, we decided in the end to refrain from including HAT-P-18 and WASP-10 in our analysis due to their cool $T_\mathrm{eff}$ values from \citetalias{brewer2018}. As discussed in \S\ref{sec:stellar_abuns}, we find good agreement between five targets that overlap between our sample and that of \citetalias{brewer2018}, but the range of $T_\mathrm{eff}$ that those targets span is limited ($\sim$5350-5800 K), and \citetalias{brewer2018} lists the $T_\mathrm{eff}$ of HAT-P-18 as 4785 K, which is cooler than the range of our comparison sample. As shown in Figure \ref{fig:gce}, the abundances of HAT-P-18 as derived in \citetalias{brewer2018} (pink circled point) place it outside the general trend of Galactic chemical evolution (as shown by the abundances of solar twins from \citealt{nissen2014}). In their analysis, Brewer \& Fischer do tend to see stronger trends with $T_\mathrm{eff}$ in their recovered abundances, which might explain the high values of HAT-P-18. However, WASP-10 is also cooler (4775 K in Brewer \& Fischer), but its [C/Fe] and [O/Fe] values seem to still lie within the general trend of Galactic chemical evolution (Figure \ref{fig:gce}, orange circled point). Furthermore, the constraints reported by \citetalias{brewer2018} on carbon are actually stronger at lower $T_\mathrm{eff}$ values due to the presence of molecular lines, and the fact that both C and O agree with each other suggests that HAT-P-18 is a genuine outlier, perhaps an alpha-enhanced metal-rich star.  In any case, this system might be special in some way. Still, some volatile vs. refractory effect remains an intriguing result that deserves further investigation as more systems are discovered that are amenable to this analysis.

Finally, we considered the relationship between $Z_p/Z_\star$ and the planet mass.  Strong correlations between these two variables were seen in \cite{miller&fortney2011} and \cite{thorngren2016}, and we are able to reproduce these results, both considering individual elements (Fig. \ref{fig:zzElement}) and the metal in aggregate (Fig. \ref{fig:volRef}, right panel).  For the aggregate metallicity, a histogram of the slope posterior reveals that zero is ruled out at about 3$\sigma$ confidence, in agreement with the Kendall's tau test.  In light of our aforementioned results, it seems likely that this relationship is largely driven by the planetary mass-metallicity relationship, rather than through a connection with the stellar abundances. Supporting this idea is the nearly identical pattern seen in Figure \ref{fig:zzElement} regardless of which element is under consideration.

\section{Summary} \label{sec:conclusions}
In this paper, we built upon the work of \cite{miller&fortney2011} and  \citeauthor{thorngren2016} (2016, \citetalias{thorngren2016}) to extend the study of the host star metallicity influence on bulk planet metallicity across a wider composition range. The first step in this study was gathering detailed host star abundances -- [C/H], [O/H], [Si/H], [Mg/H], and [Ni/H], as well as updated [Fe/H] values -- of a sample of stars hosting relatively cool giant transiting planets with well-constrained masses and radii. This survey required several semesters worth of large-telescope observations to acquire the high-resolution, high S/N data needed to derive the stellar abundances, particularly the volatiles [C/H] and [O/H], around fairly faint stars (average $V$ mag $\sim$12). We provide a thorough description of the stellar parameter and abundance determination in the first part of this paper. 

With abundances in hand, we then examined the relationship between the host star and composition and the heavy-element composition of the planet, the latter estimated via the same methods as described in \citetalias{thorngren2016}. We accounted for the correlation of abundances within a given star, a signature of Galactic chemical evolution and not specifically planet formation, and then performed a series of linear fits using a Bayesian approach that accounts for errors in both variables. We find a similar strong correlation as seen in previous works between $Z_{p}/Z_{\star}$ and $M_p$, across all elements we studied. However, given the lack of correlations observed between residual planet mass -- the heavy-element mass beyond what is expected given its total mass -- and all [Z/H] values, it seems more likely that the $Z_{p}/Z_{\star}$ vs. $M_p$ relationship is driven by the planet metallicity without much input from stellar metallicity. We find an interesting potential trend between the residual planet mass and the relative amount of volatile versus refractory material in the star, although one system (HAT-P-18) is a strong outlier to this trend. Overall, this work presents several new observational relationships between host star and planet composition that should be addressed in future theoretical studies of planet formation.

\acknowledgments
We thank the referee for their very helpful comments that improved the quality of this paper. Support for this work was provided by NASA through Hubble Fellowship grant HST-HF2-51399.001 awarded to J.K.T. by the Space Telescope Science Institute, which is operated by the Association of Universities for Research in Astronomy, Inc., for NASA, under contract NAS5-26555. We acknowledge the support of NASA XRP grant NNX16AB49G to J.J.F. This research has made use of the NASA Exoplanet Archive, which is operated by the California Institute of Technology, under contract with the National Aeronautics and Space Administration under the Exoplanet Exploration Program. This work has made use of the VALD database, operated at Uppsala University, the Institute of Astronomy RAS in Moscow, and the University of Vienna. The research shown here acknowledges use of the Hypatia Catalog Database, an online compilation of stellar abundance data as described in \citet{Hinkel2014}, which was supported by NASA's Nexus for Exoplanet System Science (NExSS) research coordination network and the Vanderbilt Initiative in Data-Intensive Astrophysics (VIDA). The authors wish to recognize and acknowledge the very significant cultural role and reverence that the summit of Maunakea has always had within the indigenous Hawaiian community.  We are most fortunate to have the opportunity to conduct observations from this mountain.

\facilities{Keck:I(HIRES),Magellan:Clay(MIKE),Exoplanet Archive}
\software{astropy \citep{astropy}, emcee \citep{emcee}}

\appendix

\section{Accounting for Potential Offsets in C and O Abundances \label{sec:offsets}}
\begin{figure*}[ht!]
    \centering
    \begin{minipage}{.51\linewidth}
    	\centering
    	\vspace{-15pt}
    	\includegraphics[width=\textwidth,clip]{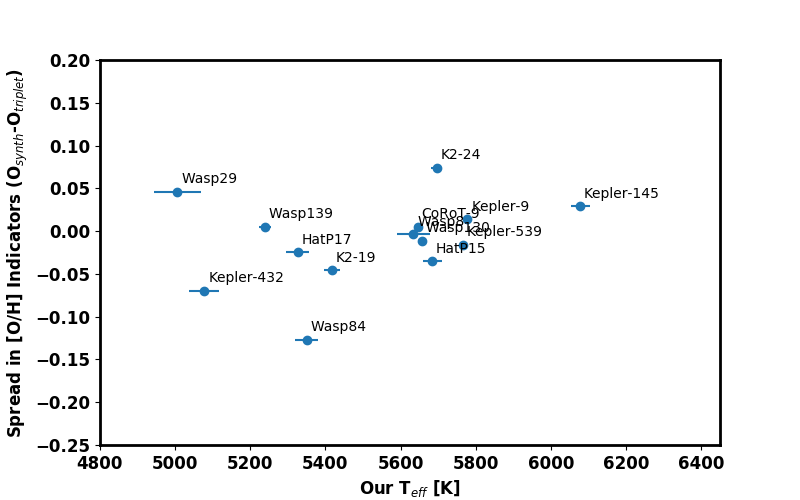} 
    \end{minipage}
    \hspace{-20pt}
    \begin{minipage}{0.51\linewidth}
    	\centering
    	\vspace{-15pt}
    	\includegraphics[width=\textwidth,clip]{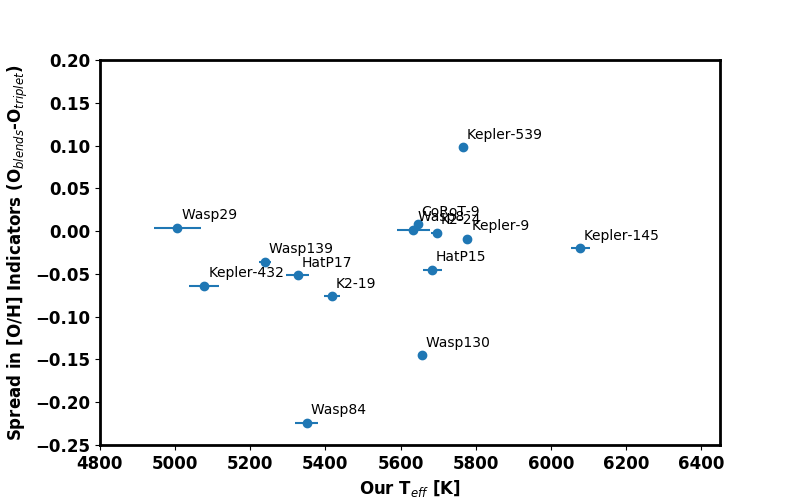}
    \end{minipage}
    \hspace{-20pt}
    \begin{minipage}{0.51\linewidth}
    	\centering
    	\vspace{-15pt}
    	\includegraphics[width=\textwidth,clip]{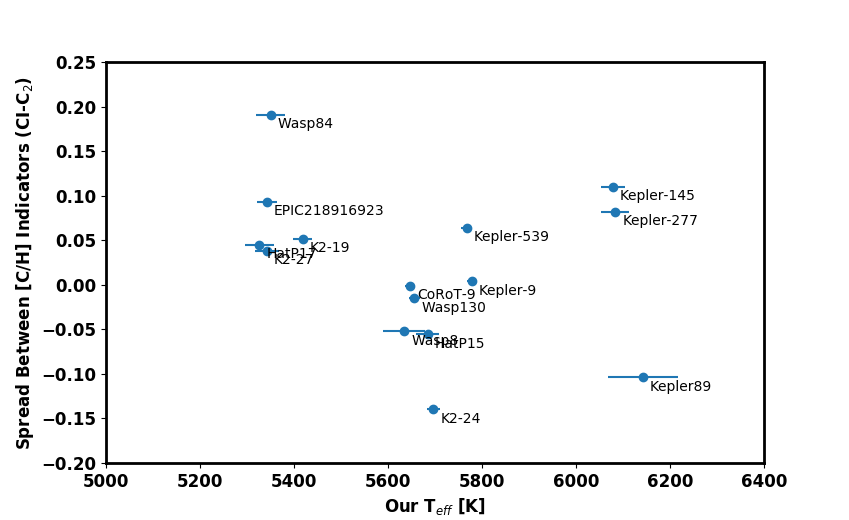}
    \end{minipage}
    \caption{These plots show the differences (spreads) in [O/H] (top) and [C/H] (bottom) abundances derived from different line indicators as a function of the effective temperature of the star as measured in this work. }
    \label{fig:empirical_corrections}
\end{figure*}

As described in \S\ref{sec:stellar_abuns}, there are several stars for which some carbon and oxygen abundance indicators were not measured or included in the calculation of the final average abundance value that is used in our analysis below. This has the potential of introducing systematic error that could influence trends in these abundances with planet parameters, and thus our conclusions. To attempt to account for potential systematic offsets, we applied an empirically derived correction to the stars with missing abundance indicators for carbon and oxygen. We took the stars in our sample for which multiple abundance indicators were measured, and calculated the difference in [C/H] or [O/H] from these indicators, [C/H]$_{\rm{spread}}$ and [O/H]$_{\rm{spread}}$, see Figure \ref{fig:empirical_corrections}. Then for the stars with missing abundance indicators -- Kepler-419, Kepler-432, WASP-139, WASP-29, Kepler-89, Kepler-419, K2-27, K2-139, Kepler-227, and HD~80606 -- we find the closest match in effective temperature to a star with multiple abundance indicators measured. If the [C/H]$_{\rm{spread}}$ or [O/H]$_{\rm{spread}}$ value is higher than the error from just one abundance indicator, we adopt the spread value as the new error. If the spread value is lower than the error from just one abundance indicator, we do not change the error. For example, in the case of Kepler-419 it has an oxygen triplet measurement but not an oxygen forbidden-line measurement. A star of comparable $T_\mathrm{eff}$ (Kepler-145) has a maximum [O/H]$_{\rm{spread}}$ (forbidden line - triplet) spread of 0.029 dex. The error on [O/H] for Kepler-419 from just the oxygen triplet is already higher than 0.029 (0.108 dex), so we do not add to this error. This is similar to the approach we took for the stars with multiple abundance indicators -- we adopt the as the final error the higher value between the formal error and the spread from different indicators.

Applying our empirically derived correction results in increased [C/H] errors for Kepler-419 and Kepler-432 and increased [O/H] errors for K2-27 and K2-139. These errors are reported in Table \ref{tab:abuns_co}. The largest increase in of error magnitude is 0.024 dex in the [C/H] value for Kepler-432, all other corrections are smaller. We adopt the corrected errors in our analysis below, although doing so does not significantly change our results from the case of using uncorrected errors. This suggests that our results are robust to small shifts in abundance values. However, it is clear from Figure \ref{fig:empirical_corrections} that [C/H]$_{\rm{spread}}$ and [O/H]$_{\rm{spread}}$ are larger below $T_{\mathrm{eff}} \sim$5500~K; inflating the errors for objects with missing abundance indicators may not fully negate systematically higher or lower abundance values, which may impact our final results. 

\section{Sensitivity Test on Final Results}

\begin{figure*}
    \includegraphics[width=\textwidth]{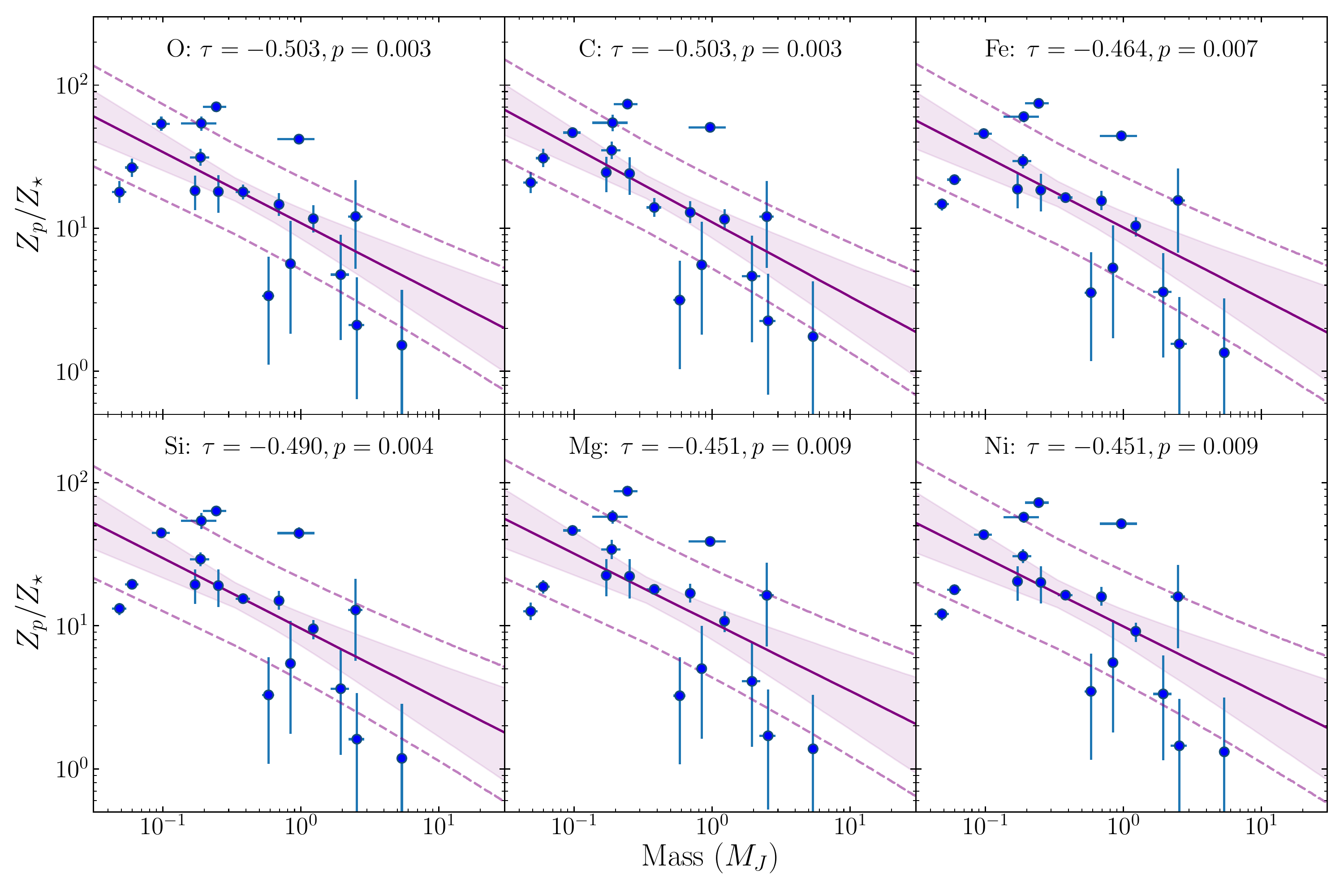}
    \caption{Same as Figure \ref{fig:zzElement}, but only including stars with [C/H] abundances derived from the C I lines only [O/H] abundances derived from the oxygen triplet.}
    \label{fig:zzElement_ST}
\end{figure*}

\begin{figure*}
    \includegraphics[width=\textwidth]{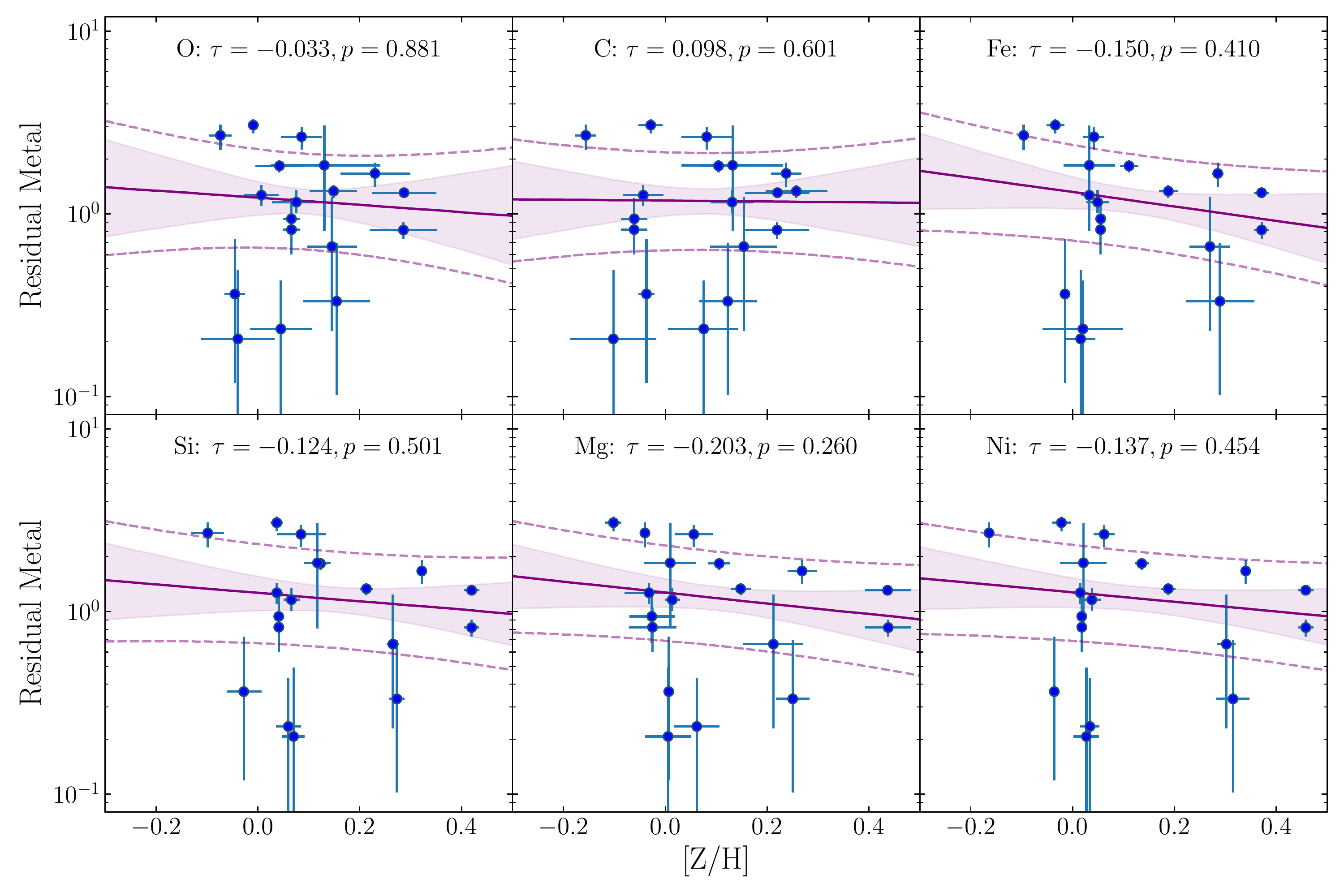}
    \caption{Same as Figure \ref{fig:residAbund}, but only including stars with [C/H] abundances derived from the C I lines only [O/H] abundances derived from the oxygen triplet.}
    \label{fig:residAbund_ST}
\end{figure*}

\begin{figure*}
    \includegraphics[width=\textwidth]{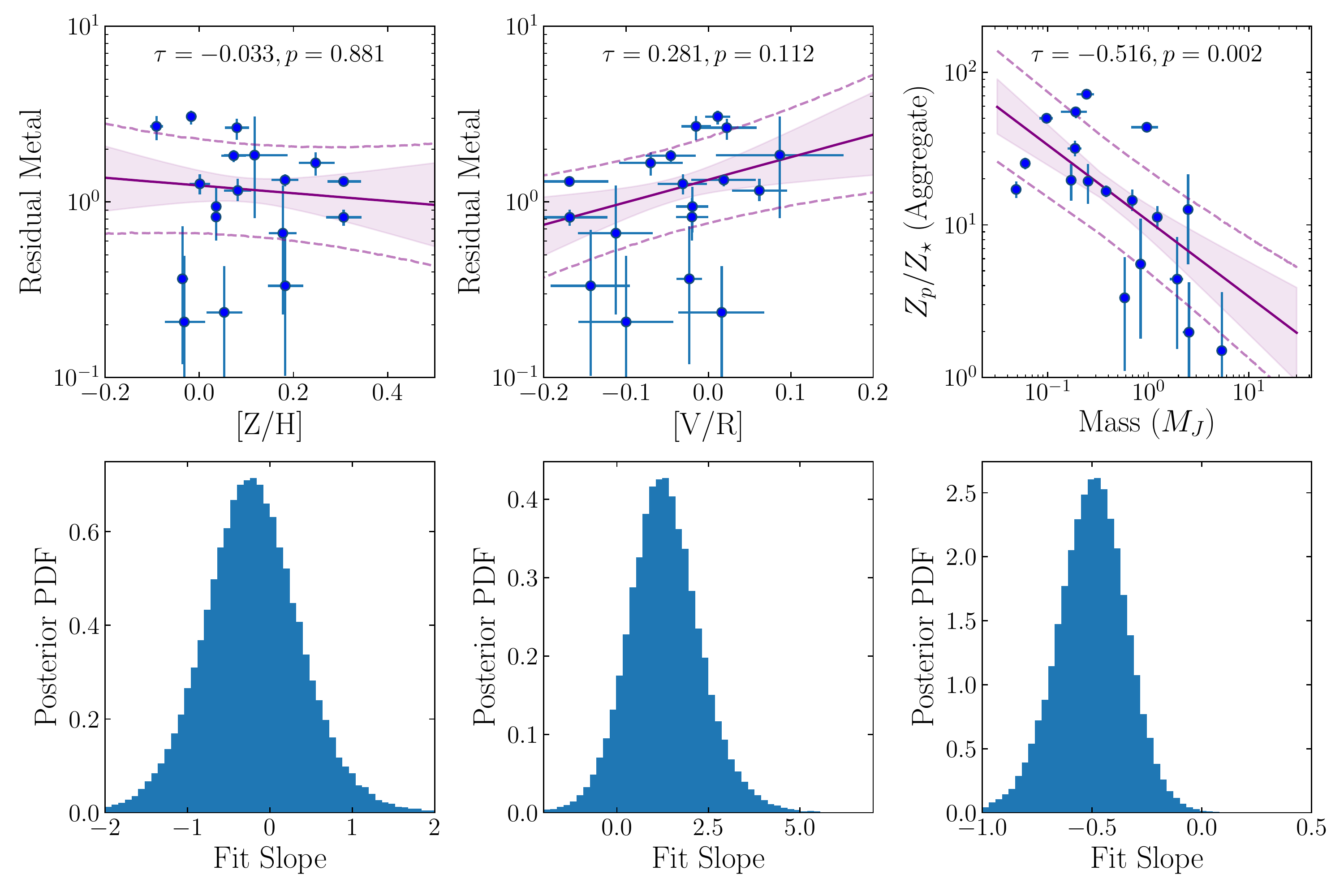}
    \caption{Same as Figure \ref{fig:volRef}, but only including stars with [C/H] abundances derived from the C I lines only [O/H] abundances derived from the oxygen triplet.}
    \label{fig:volRef_ST}
\end{figure*}

To investigate further whether the stars with missing [C/H] and/or [O/H] abundance indicators could be biasing our final results, we performed a sensitivity test by removing from the sample the stars missing the most frequently measured C and O indicators (C I and the oxygen triplet), and also removing Kepler-282 and Kepler-238 from \citetalias{brewer2018} because the authors did not report abundance results from specific lines. We then repeated our analysis with this smaller sample of stars (18), using only [C/H] derived from the C I lines and only [O/H] derived from the oxygen triplet, see Figures \ref{fig:zzElement_ST}-\ref{fig:volRef_ST}. We find that the correlations are less significant but that the removed stars are not systematically different. That is, our final conclusions do not change based on only stars with homogeneous carbon and oxygen abundance indicators alone.

\clearpage

\bibliography{references}

\begin{thebibliography}{}
\expandafter\ifx\csname natexlab\endcsname\relax\def\natexlab#1{#1}\fi

\bibitem[{Adibekyan {et~al.}(2012)Adibekyan, Santos, Sousa, Israelian,
  Delgado~Mena, Gonz\'alez~Hern\'andez, Mayor, Lovis, \& Udry}]{Adibekyan2012}
Adibekyan, V.~Z., Santos, N.~C., Sousa, S.~G., {et~al.} 2012, Astronomy and
  Astrophysics, 543, A89

\bibitem[{Ali-Dib {et~al.}(2014)Ali-Dib, Mousis, Petit, \& Lunine}]{alidib2014}
Ali-Dib, M., Mousis, O., Petit, J.-M., \& Lunine, J.~I. 2014, The Astrophysical
  Journal, 785, 125

\bibitem[{{Alibert} {et~al.}(2005){Alibert}, {Mordasini}, {Benz}, \&
  {Winisdoerffer}}]{alibert2005}
{Alibert}, Y., {Mordasini}, C., {Benz}, W., \& {Winisdoerffer}, C. 2005, \aap,
  434, 343

\bibitem[{Asplund {et~al.}(2009)Asplund, Grevesse, Sauval, \&
  Scott}]{Asplund2009}
Asplund, M., Grevesse, N., Sauval, A.~J., \& Scott, P. 2009, Annual Review of
  Astronomy and Astrophysics, 47, 481

\bibitem[{{Bedell} {et~al.}(2014){Bedell}, {Mel{\'e}ndez}, {Bean},
  {Ram{\'{\i}}rez}, {Leite}, \& {Asplund}}]{bedell2014}
{Bedell}, M., {Mel{\'e}ndez}, J., {Bean}, J.~L., {et~al.} 2014, \apj, 795, 23

\bibitem[{{Bensby} {et~al.}(2004){Bensby}, {Feltzing}, \&
  {Lundstr{\"o}m}}]{bensby2004}
{Bensby}, T., {Feltzing}, S., \& {Lundstr{\"o}m}, I. 2004, \aap, 415, 155

\bibitem[{{Bensby} {et~al.}(2014){Bensby}, {Feltzing}, \& {Oey}}]{bensby2014}
{Bensby}, T., {Feltzing}, S., \& {Oey}, M.~S. 2014, \aap, 562, A71

\bibitem[{{Bernstein} {et~al.}(2003){Bernstein}, {Shectman}, {Gunnels},
  {Mochnacki}, \& {Athey}}]{bernstein2003}
{Bernstein}, R., {Shectman}, S.~A., {Gunnels}, S.~M., {Mochnacki}, S., \&
  {Athey}, A.~E. 2003, in \procspie, Vol. 4841, Instrument Design and
  Performance for Optical/Infrared Ground-based Telescopes, ed. M.~{Iye} \&
  A.~F.~M. {Moorwood}, 1694--1704

\bibitem[{Birnstiel {et~al.}(2012)Birnstiel, Klahr, \&
  Ercolano}]{Birnstiel2012}
Birnstiel, T., Klahr, H., \& Ercolano, B. 2012, Astronomy and Astrophysics,
  539, A148

\bibitem[{{Booth} {et~al.}(2017){Booth}, {Clarke}, {Madhusudhan}, \&
  {Ilee}}]{booth2017}
{Booth}, R.~A., {Clarke}, C.~J., {Madhusudhan}, N., \& {Ilee}, J.~D. 2017,
  \mnras, 469, 3994

\bibitem[{Brauer {et~al.}(2008)Brauer, Dullemond, \& Henning}]{Brauer2008}
Brauer, F., Dullemond, C.~P., \& Henning, T. 2008, Astronomy and Astrophysics,
  480, 859

\bibitem[{{Brewer} \& {Fischer}(2018)}]{brewer2018}
{Brewer}, J.~M., \& {Fischer}, D.~A. 2018, \apjs, 237, 38

\bibitem[{{Buchhave} \& {Latham}(2015)}]{buchhave2015}
{Buchhave}, L.~A., \& {Latham}, D.~W. 2015, \apj, 808, 187

\bibitem[{{Buchhave} {et~al.}(2014){Buchhave}, {Bizzarro}, {Latham},
  {Sasselov}, {Cochran}, {Endl}, {Isaacson}, {Juncher}, \&
  {Marcy}}]{buchhave2014}
{Buchhave}, L.~A., {Bizzarro}, M., {Latham}, D.~W., {et~al.} 2014, \nat, 509,
  593

\bibitem[{{Burrows} {et~al.}(2007){Burrows}, {Hubeny}, {Budaj}, \&
  {Hubbard}}]{burrows2007}
{Burrows}, A., {Hubeny}, I., {Budaj}, J., \& {Hubbard}, W.~B. 2007, \apj, 661,
  502

\bibitem[{{Caffau} {et~al.}(2010){Caffau}, {Ludwig}, {Bonifacio}, {Faraggiana},
  {Steffen}, {Freytag}, {Kamp}, \& {Ayres}}]{caffau2010}
{Caffau}, E., {Ludwig}, H.-G., {Bonifacio}, P., {et~al.} 2010, \aap, 514, A92

\bibitem[{{Caffau} {et~al.}(2008){Caffau}, {Ludwig}, {Steffen}, {Ayres},
  {Bonifacio}, {Cayrel}, {Freytag}, \& {Plez}}]{caffau2008}
{Caffau}, E., {Ludwig}, H.-G., {Steffen}, M., {et~al.} 2008, \aap, 488, 1031

\bibitem[{{Epstein} {et~al.}(2010){Epstein}, {Johnson}, {Dong}, {Udalski},
  {Gould}, \& {Becker}}]{epstein2010}
{Epstein}, C.~R., {Johnson}, J.~A., {Dong}, S., {et~al.} 2010, \apj, 709, 447

\bibitem[{{Fabbian} {et~al.}(2009){Fabbian}, {Asplund}, {Barklem}, {Carlsson},
  \& {Kiselman}}]{fabbian2009}
{Fabbian}, D., {Asplund}, M., {Barklem}, P.~S., {Carlsson}, M., \& {Kiselman},
  D. 2009, \aap, 500, 1221

\bibitem[{Fischer \& Valenti(2005)}]{fischer&valenti2005}
Fischer, D.~A., \& Valenti, J. 2005, The Astrophysical Journal, 622, 1102

\bibitem[{{Foreman-Mackey} {et~al.}(2013){Foreman-Mackey}, {Hogg}, {Lang}, \&
  {Goodman}}]{emcee}
{Foreman-Mackey}, D., {Hogg}, D.~W., {Lang}, D., \& {Goodman}, J. 2013, \pasp,
  125, 306

\bibitem[{{Fortney} {et~al.}(2013){Fortney}, {Mordasini}, {Nettelmann},
  {Kempton}, {Greene}, \& {Zahnle}}]{fortney2013}
{Fortney}, J.~J., {Mordasini}, C., {Nettelmann}, N., {et~al.} 2013, \apj, 775,
  80

\bibitem[{Geman \& Geman(1984)}]{Geman1984}
Geman, S., \& Geman, D. 1984, IEEE Transactions on Pattern Analysis and Machine
  Intelligence, PAMI-6, 721

\bibitem[{{Gonzalez}(1997)}]{gonzalez1997}
{Gonzalez}, G. 1997, \mnras, 285, 403

\bibitem[{{Gratton} {et~al.}(1999){Gratton}, {Carretta}, {Eriksson}, \&
  {Gustafsson}}]{gratton1999}
{Gratton}, R.~G., {Carretta}, E., {Eriksson}, K., \& {Gustafsson}, B. 1999,
  \aap, 350, 955

\bibitem[{{Guillot} {et~al.}(2006){Guillot}, {Santos}, {Pont}, {Iro}, {Melo},
  \& {Ribas}}]{guillot2006}
{Guillot}, T., {Santos}, N.~C., {Pont}, F., {et~al.} 2006, \aap, 453, L21

\bibitem[{{Gustafsson} {et~al.}(2008){Gustafsson}, {Edvardsson}, {Eriksson},
  {J{\o}rgensen}, {Nordlund}, \& {Plez}}]{gustafsson2008}
{Gustafsson}, B., {Edvardsson}, B., {Eriksson}, K., {et~al.} 2008, \aap, 486,
  951

\bibitem[{{Hasegawa} {et~al.}(2018){Hasegawa}, {Bryden}, {Ikoma}, {Vasisht}, \&
  {Swain}}]{hasegawa2018}
{Hasegawa}, Y., {Bryden}, G., {Ikoma}, M., {Vasisht}, G., \& {Swain}, M. 2018,
  \apj, 865, 32

\bibitem[{Hastings(1970)}]{Hastings1970}
Hastings, W.~K. 1970, Biometrika, 57, 97

\bibitem[{Haywood(2009)}]{Haywood2009}
Haywood, M. 2009, The Astrophysical Journal Letters, 698, L1

\bibitem[{{Helled} \& {Stevenson}(2017)}]{helled2017}
{Helled}, R., \& {Stevenson}, D. 2017, \apjl, 840, L4

\bibitem[{Hinkel {et~al.}(2014)Hinkel, Timmes, Young, Pagano, \&
  Turnbull}]{Hinkel2014}
Hinkel, N.~R., Timmes, F.~X., Young, P.~A., Pagano, M.~D., \& Turnbull, M.~C.
  2014, The Astronomical Journal, 148, 54

\bibitem[{Hinkel {et~al.}(2017)Hinkel, Mamajek, Turnbull, Osby, Shkolnik,
  Smith, Klimasewski, Somers, \& Desch}]{Hinkel2017}
Hinkel, N.~R., Mamajek, E.~E., Turnbull, M.~C., {et~al.} 2017, The
  Astrophysical Journal, 848, 34

\bibitem[{{Johnson} {et~al.}(2010){Johnson}, {Aller}, {Howard}, \&
  {Crepp}}]{johnson2010}
{Johnson}, J.~A., {Aller}, K.~M., {Howard}, A.~W., \& {Crepp}, J.~R. 2010,
  \pasp, 122, 905

\bibitem[{{Kelson}(2003)}]{Kelson2003}
{Kelson}, D.~D. 2003, \pasp, 115, 688

\bibitem[{{Kelson} {et~al.}(2000){Kelson}, {Illingworth}, {van Dokkum}, \&
  {Franx}}]{Kelson2000}
{Kelson}, D.~D., {Illingworth}, G.~D., {van Dokkum}, P.~G., \& {Franx}, M.
  2000, \apj, 531, 159

\bibitem[{{Kiselman}(1991)}]{kiselman1991}
{Kiselman}, D. 1991, \aap, 245, L9

\bibitem[{{Kiselman}(1993)}]{kiselman1993}
---. 1993, \aap, 275

\bibitem[{{Konopacky} {et~al.}(2013){Konopacky}, {Barman}, {Macintosh}, \&
  {Marois}}]{konopacky2013}
{Konopacky}, Q.~M., {Barman}, T.~S., {Macintosh}, B.~A., \& {Marois}, C. 2013,
  Science, 339, 1398

\bibitem[{{Kreidberg} {et~al.}(2014){Kreidberg}, {Bean}, {D{\'e}sert}, {Line},
  {Fortney}, {Madhusudhan}, {Stevenson}, {Showman}, {Charbonneau},
  {McCullough}, {Seager}, {Burrows}, {Henry}, {Williamson}, {Kataria}, \&
  {Homeier}}]{kreidberg2014}
{Kreidberg}, L., {Bean}, J.~L., {D{\'e}sert}, J.-M., {et~al.} 2014, \apjl, 793,
  L27

\bibitem[{{Kupka} {et~al.}(1999){Kupka}, {Piskunov}, {Ryabchikova}, {Stempels},
  \& {Weiss}}]{kupka1999}
{Kupka}, F., {Piskunov}, N., {Ryabchikova}, T.~A., {Stempels}, H.~C., \&
  {Weiss}, W.~W. 1999, \aaps, 138, 119

\bibitem[{{Kupka} {et~al.}(2000){Kupka}, {Ryabchikova}, {Piskunov}, {Stempels},
  \& {Weiss}}]{kupka2000}
{Kupka}, F.~G., {Ryabchikova}, T.~A., {Piskunov}, N.~E., {Stempels}, H.~C., \&
  {Weiss}, W.~W. 2000, Baltic Astronomy, 9, 590

\bibitem[{{Lambert} \& {Ries}(1981)}]{lambert&ries1981}
{Lambert}, D.~L., \& {Ries}, L.~M. 1981, \apj, 248, 228

\bibitem[{{Madhusudhan} {et~al.}(2011){Madhusudhan}, {Mousis}, {Johnson}, \&
  {Lunine}}]{madhusudhan2011}
{Madhusudhan}, N., {Mousis}, O., {Johnson}, T.~V., \& {Lunine}, J.~I. 2011,
  \apj, 743, 191

\bibitem[{{Marley} {et~al.}(2007){Marley}, {Fortney}, {Hubickyj},
  {Bodenheimer}, \& {Lissauer}}]{marley2007}
{Marley}, M.~S., {Fortney}, J.~J., {Hubickyj}, O., {Bodenheimer}, P., \&
  {Lissauer}, J.~J. 2007, \apj, 655, 541

\bibitem[{Miller \& Fortney(2011)}]{miller&fortney2011}
Miller, N., \& Fortney, J.~J. 2011, The Astrophysical Journal Letters, 736, L29

\bibitem[{{Mordasini}(2014)}]{mordasini2014b}
{Mordasini}, C. 2014, \aap, 572, A118

\bibitem[{{Mordasini} {et~al.}(2008){Mordasini}, {Alibert}, {Benz}, \&
  {Naef}}]{Mordasini2008}
{Mordasini}, C., {Alibert}, Y., {Benz}, W., \& {Naef}, D. 2008, in Astronomical
  Society of the Pacific Conference Series, Vol. 398, Extreme Solar Systems,
  ed. D.~{Fischer}, F.~A. {Rasio}, S.~E. {Thorsett}, \& A.~{Wolszczan}, 235

\bibitem[{{Mordasini} {et~al.}(2009){Mordasini}, {Alibert}, {Benz}, \&
  {Naef}}]{mordasini2009}
{Mordasini}, C., {Alibert}, Y., {Benz}, W., \& {Naef}, D. 2009, \aap, 501, 1161

\bibitem[{Mordasini {et~al.}(2014)Mordasini, Klahr, Alibert, Miller, \&
  Henning}]{Mordasini2014a}
Mordasini, C., Klahr, H., Alibert, Y., Miller, N., \& Henning, T. 2014,
  Astronomy \& Astrophysics, 566, A141

\bibitem[{{Mordasini} {et~al.}(2016){Mordasini}, {van Boekel}, {Molli{\`e}re},
  {Henning}, \& {Benneke}}]{mordasini2016}
{Mordasini}, C., {van Boekel}, R., {Molli{\`e}re}, P., {Henning}, T., \&
  {Benneke}, B. 2016, \apj, 832, 41

\bibitem[{{Mortier} {et~al.}(2013){Mortier}, {Santos}, {Sousa}, {Adibekyan},
  {Delgado Mena}, {Tsantaki}, {Israelian}, \& {Mayor}}]{mortier2013}
{Mortier}, A., {Santos}, N.~C., {Sousa}, S.~G., {et~al.} 2013, \aap, 557, A70

\bibitem[{{Mulders} {et~al.}(2016){Mulders}, {Pascucci}, {Apai}, {Frasca}, \&
  {Molenda-{\.Z}akowicz}}]{mulders2016}
{Mulders}, G.~D., {Pascucci}, I., {Apai}, D., {Frasca}, A., \&
  {Molenda-{\.Z}akowicz}, J. 2016, \aj, 152, 187

\bibitem[{{Nissen} {et~al.}(2014){Nissen}, {Chen}, {Carigi}, {Schuster}, \&
  {Zhao}}]{nissen2014}
{Nissen}, P.~E., {Chen}, Y.~Q., {Carigi}, L., {Schuster}, W.~J., \& {Zhao}, G.
  2014, \aap, 568, A25

\bibitem[{{Noguchi} {et~al.}(2002){Noguchi}, {Aoki}, {Kawanomoto}, {Ando},
  {Honda}, {Izumiura}, {Kambe}, {Okita}, {Sadakane}, {Sato}, {Tajitsu},
  {Takada-Hidai}, {Tanaka}, {Watanabe}, \& {Yoshida}}]{Noguchi2002}
{Noguchi}, K., {Aoki}, W., {Kawanomoto}, S., {et~al.} 2002, \pasj, 54, 855

\bibitem[{{Petigura} {et~al.}(2018){Petigura}, {Marcy}, {Winn}, {Weiss},
  {Fulton}, {Howard}, {Sinukoff}, {Isaacson}, {Morton}, \&
  {Johnson}}]{petigura2018}
{Petigura}, E.~A., {Marcy}, G.~W., {Winn}, J.~N., {et~al.} 2018, \aj, 155, 89

\bibitem[{{Piskunov} \& {Valenti}(2017)}]{Piskunov&Valenti2017}
{Piskunov}, N., \& {Valenti}, J.~A. 2017, \aap, 597, A16

\bibitem[{{Piskunov} {et~al.}(1995){Piskunov}, {Kupka}, {Ryabchikova}, {Weiss},
  \& {Jeffery}}]{piskunov1995}
{Piskunov}, N.~E., {Kupka}, F., {Ryabchikova}, T.~A., {Weiss}, W.~W., \&
  {Jeffery}, C.~S. 1995, \aaps, 112, 525

\bibitem[{{Pollack} {et~al.}(1996){Pollack}, {Hubickyj}, {Bodenheimer},
  {Lissauer}, {Podolak}, \& {Greenzweig}}]{pollack1996}
{Pollack}, J.~B., {Hubickyj}, O., {Bodenheimer}, P., {et~al.} 1996, \icarus,
  124, 62

\bibitem[{Powell {et~al.}(2017)Powell, {Murray-Clay}, \&
  Schlichting}]{Powell2017}
Powell, D., {Murray-Clay}, R., \& Schlichting, H.~E. 2017, The Astrophysical
  Journal, 840, 93

\bibitem[{{Ram{\'{\i}}rez} {et~al.}(2007){Ram{\'{\i}}rez}, {Allende Prieto}, \&
  {Lambert}}]{ramirez2007}
{Ram{\'{\i}}rez}, I., {Allende Prieto}, C., \& {Lambert}, D.~L. 2007, \aap,
  465, 271

\bibitem[{{Ram{\'{\i}}rez} {et~al.}(2014){Ram{\'{\i}}rez}, {Mel{\'e}ndez},
  {Bean}, {Asplund}, {Bedell}, {Monroe}, {Casagrande}, {Schirbel}, {Dreizler},
  {Teske}, {Tucci Maia}, {Alves-Brito}, \& {Baumann}}]{ramirez2014}
{Ram{\'{\i}}rez}, I., {Mel{\'e}ndez}, J., {Bean}, J., {et~al.} 2014, \aap, 572,
  A48

\bibitem[{Raymond {et~al.}(2006)Raymond, Mandell, \& Sigurdsson}]{raymond2006}
Raymond, S.~N., Mandell, A.~M., \& Sigurdsson, S. 2006, Science, 313, 1413

\bibitem[{{Ryabchikova} {et~al.}(2015){Ryabchikova}, {Piskunov}, {Kurucz},
  {Stempels}, {Heiter}, {Pakhomov}, \& {Barklem}}]{ryabchikova2015}
{Ryabchikova}, T., {Piskunov}, N., {Kurucz}, R.~L., {et~al.} 2015, \physscr,
  90, 054005

\bibitem[{{Ryabchikova} {et~al.}(1997){Ryabchikova}, {Piskunov}, {Kupka}, \&
  {Weiss}}]{ryabchikova1997}
{Ryabchikova}, T.~A., {Piskunov}, N.~E., {Kupka}, F., \& {Weiss}, W.~W. 1997,
  Baltic Astronomy, 6, 244

\bibitem[{{Santos} {et~al.}(2004){Santos}, {Israelian}, \&
  {Mayor}}]{santos2004}
{Santos}, N.~C., {Israelian}, G., \& {Mayor}, M. 2004, \aap, 415, 1153

\bibitem[{{Schlaufman}(2015)}]{schlaufman2015}
{Schlaufman}, K.~C. 2015, \apjl, 799, L26

\bibitem[{{Schuler} {et~al.}(2011){Schuler}, {Flateau}, {Cunha}, {King},
  {Ghezzi}, \& {Smith}}]{schuler2011}
{Schuler}, S.~C., {Flateau}, D., {Cunha}, K., {et~al.} 2011, \apj, 732, 55

\bibitem[{{Sneden}(1973)}]{sneden1973}
{Sneden}, C. 1973, \apj, 184, 839

\bibitem[{{Storey} \& {Zeippen}(2000)}]{storey&zeippen2000}
{Storey}, P.~J., \& {Zeippen}, C.~J. 2000, \mnras, 312, 813

\bibitem[{{Takeda}(2003)}]{takeda2003}
{Takeda}, Y. 2003, \aap, 402, 343

\bibitem[{{Teske} {et~al.}(2013){Teske}, {Cunha}, {Schuler}, {Griffith}, \&
  {Smith}}]{teske2013}
{Teske}, J.~K., {Cunha}, K., {Schuler}, S.~C., {Griffith}, C.~A., \& {Smith},
  V.~V. 2013, \apj, 778, 132

\bibitem[{{Teske} {et~al.}(2014){Teske}, {Cunha}, {Smith}, {Schuler}, \&
  {Griffith}}]{teske2014}
{Teske}, J.~K., {Cunha}, K., {Smith}, V.~V., {Schuler}, S.~C., \& {Griffith},
  C.~A. 2014, \apj, 788, 39

\bibitem[{{The Astropy Collaboration} {et~al.}(2018){The Astropy
  Collaboration}, {Price-Whelan}, {Sip{\H o}cz}, {G{\"u}nther}, {Lim},
  {Crawford}, {Conseil}, {Shupe}, {Craig}, {Dencheva}, {Ginsburg},
  {VanderPlas}, {Bradley}, {P{\'e}rez-Su{\'a}rez}, {de Val-Borro}, {Aldcroft},
  {Cruz}, {Robitaille}, {Tollerud}, {Ardelean}, {Babej}, {Bachetti}, {Bakanov},
  {Bamford}, {Barentsen}, {Barmby}, {Baumbach}, {Berry}, {Biscani}, {Boquien},
  {Bostroem}, {Bouma}, {Brammer}, {Bray}, {Breytenbach}, {Buddelmeijer},
  {Burke}, {Calderone}, {Cano Rodr{\'{\i}}guez}, {Cara}, {Cardoso},
  {Cheedella}, {Copin}, {Crichton}, {D{\'A}vella}, {Deil}, {Depagne},
  {Dietrich}, {Donath}, {Droettboom}, {Earl}, {Erben}, {Fabbro}, {Ferreira},
  {Finethy}, {Fox}, {Garrison}, {Gibbons}, {Goldstein}, {Gommers}, {Greco},
  {Greenfield}, {Groener}, {Grollier}, {Hagen}, {Hirst}, {Homeier}, {Horton},
  {Hosseinzadeh}, {Hu}, {Hunkeler}, {Ivezi{\'c}}, {Jain}, {Jenness}, {Kanarek},
  {Kendrew}, {Kern}, {Kerzendorf}, {Khvalko}, {King}, {Kirkby}, {Kulkarni},
  {Kumar}, {Lee}, {Lenz}, {Littlefair}, {Ma}, {Macleod}, {Mastropietro},
  {McCully}, {Montagnac}, {Morris}, {Mueller}, {Mumford}, {Muna}, {Murphy},
  {Nelson}, {Nguyen}, {Ninan}, {N{\"o}the}, {Ogaz}, {Oh}, {Parejko}, {Parley},
  {Pascual}, {Patil}, {Patil}, {Plunkett}, {Prochaska}, {Rastogi}, {Reddy
  Janga}, {Sabater}, {Sakurikar}, {Seifert}, {Sherbert}, {Sherwood-Taylor},
  {Shih}, {Sick}, {Silbiger}, {Singanamalla}, {Singer}, {Sladen}, {Sooley},
  {Sornarajah}, {Streicher}, {Teuben}, {Thomas}, {Tremblay}, {Turner},
  {Terr{\'o}n}, {van Kerkwijk}, {de la Vega}, {Watkins}, {Weaver}, {Whitmore},
  {Woillez}, \& {Zabalza}}]{astropy}
{The Astropy Collaboration}, {Price-Whelan}, A.~M., {Sip{\H o}cz}, B.~M.,
  {et~al.} 2018, ArXiv e-prints, arXiv:1801.02634

\bibitem[{{Thevenin}(1990)}]{thevenin1990}
{Thevenin}, F. 1990, \aaps, 82, 179

\bibitem[{Thorngren {et~al.}(2016)Thorngren, Fortney, Murray-Clay, \&
  Lopez}]{thorngren2016}
Thorngren, D.~P., Fortney, J.~J., Murray-Clay, R.~A., \& Lopez, E.~D. 2016, The
  Astrophysical Journal, 831, 64

\bibitem[{{Urdahl} {et~al.}(1991){Urdahl}, {Bao}, \& {Jackson}}]{urdahl1991}
{Urdahl}, R.~S., {Bao}, Y., \& {Jackson}, W.~M. 1991, Chemical Physics Letters,
  178, 425

\bibitem[{{Vazan} {et~al.}(2016){Vazan}, {Helled}, {Podolak}, \&
  {Kovetz}}]{vazan2016}
{Vazan}, A., {Helled}, R., {Podolak}, M., \& {Kovetz}, A. 2016, \apj, 829, 118

\bibitem[{{Vogt} {et~al.}(1994){Vogt}, {Allen}, {Bigelow}, {Bresee}, {Brown},
  {Cantrall}, {Conrad}, {Couture}, {Delaney}, {Epps}, {Hilyard}, {Hilyard},
  {Horn}, {Jern}, {Kanto}, {Keane}, {Kibrick}, {Lewis}, {Osborne},
  {Pardeilhan}, {Pfister}, {Ricketts}, {Robinson}, {Stover}, {Tucker}, {Ward},
  \& {Wei}}]{Vogt1994}
{Vogt}, S.~S., {Allen}, S.~L., {Bigelow}, B.~C., {et~al.} 1994, in \procspie,
  Vol. 2198, Instrumentation in Astronomy VIII, ed. D.~L. {Crawford} \& E.~R.
  {Craine}, 362

\bibitem[{{Wakeford} {et~al.}(2017){Wakeford}, {Sing}, {Kataria}, {Deming},
  {Nikolov}, {Lopez}, {Tremblin}, {Amundsen}, {Lewis}, {Mandell}, {Fortney},
  {Knutson}, {Benneke}, \& {Evans}}]{wakeford2017}
{Wakeford}, H.~R., {Sing}, D.~K., {Kataria}, T., {et~al.} 2017, Science, 356,
  628

\bibitem[{{Wang} \& {Fischer}(2015)}]{wang&fischer2015}
{Wang}, J., \& {Fischer}, D.~A. 2015, \aj, 149, 14

\bibitem[{{Wilson} {et~al.}(2018){Wilson}, {Teske}, {Majewski}, {Cunha},
  {Smith}, {Souto}, {Bender}, {Mahadevan}, {Troup}, {Allende Prieto},
  {Stassun}, {Skrutskie}, {Almeida}, {Garc{\'{\i}}a-Hern{\'a}ndez}, {Zamora},
  \& {Brinkmann}}]{wilson2018}
{Wilson}, R.~F., {Teske}, J., {Majewski}, S.~R., {et~al.} 2018, \aj, 155, 68

\bibitem[{{Winn} {et~al.}(2017){Winn}, {Sanchis-Ojeda}, {Rogers}, {Petigura},
  {Howard}, {Isaacson}, {Marcy}, {Schlaufman}, {Cargile}, \& {Hebb}}]{winn2017}
{Winn}, J.~N., {Sanchis-Ojeda}, R., {Rogers}, L., {et~al.} 2017, \aj, 154, 60

\bibitem[{{Yang} {et~al.}(2018){Yang}, {Mac Low}, \& {Johansen}}]{Yang2018}
{Yang}, C.-C., {Mac Low}, M.-M., \& {Johansen}, A. 2018, \apj, 868, 27

\bibitem[{{Zhu} {et~al.}(2016){Zhu}, {Wang}, \& {Huang}}]{zhu2016}
{Zhu}, W., {Wang}, J., \& {Huang}, C. 2016, \apj, 832, 196

\end{thebibliography}
\end{document}